\documentstyle[epsfig,12pt]{article}
\def\CsI{CsI(Tl)}
\def\degc{\mbox{$^\circ$C}}
\def\emi{\mbox{\,e$_0$}}

\def\k{\mbox{$^{40}$K}}
\def\na{\mbox{$^{22}$Na}}
\def\co{\mbox{$^{60}$Co}}
\def\y{\mbox{$^{88}$Y}}
\def\cs{\mbox{$^{137}$Cs}}
\def\am{\mbox{$^{241}$Am}}
\def\ly{light yield}
\def\LY{\mbox{LY}}
\def\py{photoelectron yield}
\def\PY{\mbox{PY}}

\def\enc{equivalent noise charge}
\def\ENC{\mbox{ENC}}
\def\ene{equivalent noise energy}
\def\ENE{\mbox{ENE}}
\newcommand{\mus}          {\mbox{$\,\mu\mbox{s}$}}
\newcommand{\m}            {\mbox{$\,\mbox{min}$}}
\newcommand{\nm}           {\mbox{$\,\mbox{nm}$}}
\newcommand{\mum}          {\mbox{$\,\mu\mbox{m}$}}
\newcommand{\mm}           {\mbox{$\,\mbox{mm}$}}
\newcommand{\cm}           {\mbox{$\,\mbox{cm}$}}
\newcommand{\smm}          {\mbox{$\,\mbox{mm}^2$}}
\newcommand{\scm}          {\mbox{$\,\mbox{cm}^2$}}
\newcommand{\eV}           {\mbox{$\,\mbox{eV}$}}
\newcommand{\keV}          {\mbox{$\,\mbox{keV}$}}
\newcommand{\MeV}          {\mbox{$\,\mbox{MeV}$}}
\newcommand{\V}            {\mbox{$\,\mbox{V}$}}
\newcommand{\kV}           {\mbox{$\,\mbox{kV}$}}
\newcommand{\muA}          {\mbox{$\,\mu\mbox{A}$}}
\newcommand{\nA}           {\mbox{$\,\mbox{nA}$}}
\newcommand{\pF}           {\mbox{$\,\mbox{pF}$}}
\newcommand{\nF}           {\mbox{$\,\mbox{nF}$}}
\newcommand{\MOhm}         {\mbox{$\,\mbox{M}\Omega$}}
\newcommand{\K}            {\mbox{$\,\mbox{K}$}}
\newcommand{\rad}	   {\mbox{$\,\mbox{rad}$}}
\newcommand{\krad}	   {\mbox{$\,\mbox{krad}$}}

\begin{document}

\begin{titlepage}

\title
{\bf Properties of \CsI\ Crystals and their Optimization for Calorimetry of 
High Energy Photons
\thanks{Work supported by BMBF under contract No. 06~DD~558~I}}

\author{
J. Brose\thanks{Corresponding author. E-mail: J.Brose@physik.tu-dresden.de, Fax. +49\,\,351\,463\,7292}, G. Dahlinger, K. R. Schubert \\ 
Institut f\"ur Kern- und Teilchenphysik \\ 
TU Dresden \\
D 01062 Dresden, Germany}

\date{\today}

\thispagestyle{empty}
\maketitle

\vspace*{-11cm}
\begin{flushright}
{\bf TUD-IKTP/98-01}
\end{flushright}
\vspace*{11cm}

\begin{abstract}
\it
A photomultiplier setup for precise relative \CsI\ crystal \ly\ and uniformity
measurements is described. 
It is used for wrapping material studies to optimize the uniformity and the
yield of the light output of $36\cm$ long crystals. 
The uniformity is an important property in high energy photon
calorimetry.
Results of an optimization of photodiode coupling to crystals, the influence
of temperature and radiation damage to light and \py\ are also presented.

\end{abstract}

\thispagestyle{empty}
\end{titlepage}

%%%%%%%%%%%%%%%%%%%%%%%%%%%%%%%%%%%%%%%%%%%%%%%%%%%%%%%%%%%%%%%%%%%%%%%%%%%%%
%%%%%%%%%%%%%%%%%%%%%%%%%%%%%%%%%%%%%%%%%%%%%%%%%%%%%%%%%%%%%%%%%%%%%%%%%%%%%

\section{Introduction}
\label{sec:intro}

Although Thallium doped CsI crystals are widely used in high energy physics
detectors \cite{ref:cleo,ref:cb}, new precision experiments at the
B-meson factories presently under construction~\cite{ref:babar,ref:belle}
rely on \CsI\ calorimeters with improved energy resolution, electronic noise, 
and crystal radiation hardness. The energy resolution at low energies is 
influenced by the (temperature dependent) crystal light yield, which should be
stable over long time periods and maximized to achieve low electronic 
noise~\cite{ref:babar}. 
Because of fluctuations of electromagnetic showers in the crystals it is 
necessary to know and to optimize the uniformity of light output. Related 
studies can be found in the literature 
\cite{ref:grassmann,ref:blucher,ref:valentine}, but there are often 
descrepancies between laboratory studies performed with photomultipliers 
and the actual photodiode readout used in calorimeters. One reason is the 
very different sensitivity of both readout devices at the wavelength of the 
\CsI\ scintillation light.\\

We will limit ourselves to investigations which are relevant for large high
energy physics calorimeters. After introducing setups, which allow comparable
and precise crystal measurements with photomultipliers and photodiodes, 
we describe a method to improve the crystal light yield and response 
uniformity.
Using light yield optimized crystals we studied several ways to couple 
photodiodes to crystals and to enhance the \py\, thereby minimizing the 
equivalent electronic noise energy. Since temperature and 
radiation can alter the crystal light yield, their influence was investigated 
with both readout setups. These investigations were performed as part of 
early studies for the electromagnetic calorimeter of the BABAR detector.

%%%%%%%%%%%%%%%%%%%%%%%%%%%%%%%%%%%%%%%%%%%%%%%%%%%%%%%%%%%%%%%%%%%%%%%%%%%%%

\section{Experimental Setup}
\label{sec:setup}

The \CsI\ crystals used in this study were supplied by Monokristal/Khar\-kov (Ukraine). Their Thallium content amounts to $\sim 0.08\, \mbox{mol}\%$. The crystals were slightly tapered with a front face of $6\times 5\scm$, rear face dimensions of $6\times 6 \scm$, and a length of $36\cm$, which corresponds to 19.4 radiation lengths.\\

\newpage

\subsection{Photomultiplier Readout}
\label{sec:pm}

The scintillation light of \CsI\ with a peak emission wavelength at 
$\sim 550\nm$ was read out using a Hamamatsu R669 photomultiplier tube with a 
multialkali photocathode ($5\cm$ diameter), which has an enhanced sensitivity 
in this wavelength range~\cite{ref:hamamatsupm, ref:bn206}. 
The photomultiplier, which was coupled to the crystals via a minimal airgap, was operated at $1\kV$ (anode grounded), and read out by AC coupling to an Ortec preamplifier model 113 with an input capacity of $1\nF$. A spectroscopic amplifier, Ortec model 672, provided a Gaussian shaping with $1\mus$ time constant and an additional amplification. A peak sensitive ADC, Ortec 800, was DC coupled to the amplifier in order to avoid an additional differentiation of the pulse at the ADC input. For pile-up rejection the corresponding signal from the shaping amplifier was fed into the ADC. A PC based multichannel analyzer from Target with the corresponding TMCA software was used for data acquisition.\\

\begin{figure}[t]
\begin{center}
\mbox{\epsfig{file=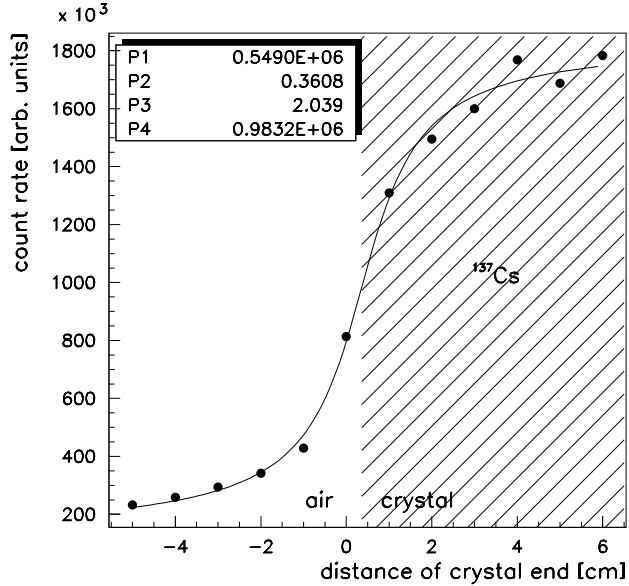,width=0.6\textwidth}}
\caption[]{Photopeak rate for $662\keV$ photons (\cs ) vs. collimator position near the crystal edge. The dashed area corresponds to a collimator position above the crystal, otherwise it is above the surrounding air. Parameter P3 is the mean beam spot diameter (in \cm ) as result of a fit to the data points.}
\label{fig:coll} 
\end{center}
\end{figure}

For position dependent measurements of the light output, a radioactive $\gamma$ source with a lead collimator was moved along the length of the crystal. The energy of the $\gamma$ source, the collimator outer dimensions, and the pin-hole diameter of the collimator were optimized in order to get a focused illumination at the crystal surface and a sufficient penetration depth. Finally the collimator was laid out as an almost half-sphere with $5\cm$ radius and a pin-hole diameter of $6\mm$. It was located $1\cm$ above the crystal surface and contained
a \cs\ source with $1\,\mu\mbox{Ci}$ and $662\keV$ $\gamma$ energy. These parameters resulted in a mean spot diameter at the crystal surface of $\sim 2\cm$ as shown in Fig.~\ref{fig:coll}.\\

A step-width of $2\cm$ was chosen for the motorized translation stage moving the collimator because of the spot size. At each collimator position a dead-time corrected spectrum was taken for 100 seconds, resulting in a total measurement time of about half an hour for 18 positions along the crystal. The region of the photopeak was fit to the sum of a Gaussian distribution and an exponential background.
In order to quote a crystal light yield, the 18 peak positions were averaged and related to the position of the corresponding peak for a small cubic \CsI\ crystal ($25\mm$ on an edge).  This  standard crystal was remeasured regularly. Thus, total \ly\ \LY\ values given in this article are relative to this standard crystal.\\

\begin{figure}[t]
\begin{center}
\setlength{\unitlength}{1mm}
\begin{picture}(80,80)
\put(58,5){\makebox{ADC channel}}
\put(-5,86){\makebox{\#}}
\mbox{\epsfig{file=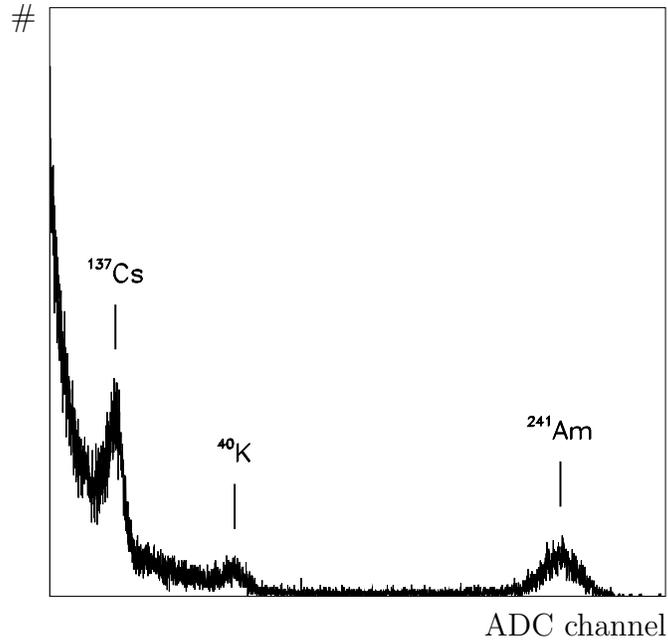,width=0.6\textwidth}}
\end{picture}
\caption[]{Pulse height spectrum for a \cs\ $\gamma$ -source with \am\ as light reference. Also visible is a \k -line from laboratory wall material.}
\label{fig:csam} 
\end{center}
\end{figure}

For precise measurements of different crystals, independent of photomultiplier drift, gain variations, and environmental influences, two different monitoring systems were developed.
One consisted of an \am\ source placed on a small hole in the crystal wrapping material at the front end of the crystal. This source emits $\alpha$-particles with an energy of about $5.4\MeV$, which generate a constant amount of scintillation light in the crystal. Fig.~\ref{fig:csam} shows a pulse height spectrum of a $^{137}$Cs $\gamma$-source with a $^{241}$Am $\alpha$-source as reference attached to a large crystal. When the crystal is scanned with the
$^{137}$Cs $\gamma$-source, the variations of the position of the $\alpha$-peak indicate drift of the electronics during the uniformity measurement, which is corrected during the analysis~\cite{ref:bn175}.
For this monitoring method, the error of the peak position was $\pm 0.3\%$ for spectra taken along one crystal, mainly due to the description of the background. However, the error for the total light yield of a particular crystal is much larger, since the $\alpha$-line position may vary from crystal to crystal owing to variations in light generation and transmission. Therefore, an external reference system was developed and operated under stable and controlled conditions.\\

\begin{figure}[t]
\begin{center}
\mbox{\epsfig{file=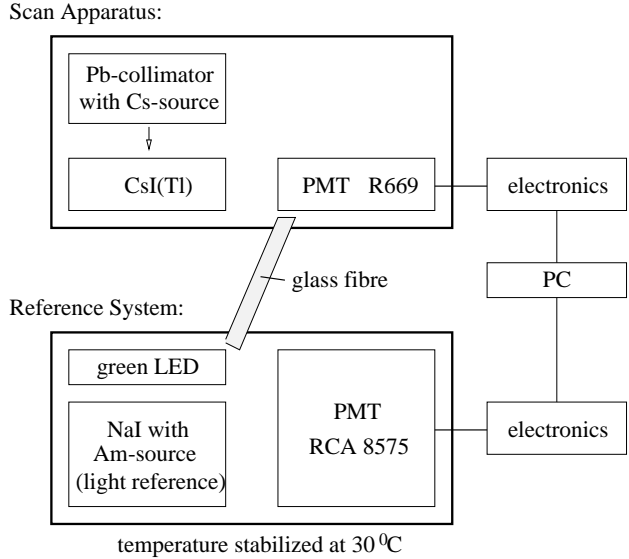,width=0.6\textwidth}}
\caption[]{Crystal scan apparatus and external reference system.}
\label{fig:refsetup} 
\end{center}
\end{figure}

This system used a ``light normal'' consisting of an encapsulated combination of a NaI(Tl) crystal and a \am\ source. Since the light output of this crystal is low, its light cannot directly be coupled to the R669 photomultiplier. Instead, it is detected by another photomultiplier (Philips RCA 8575), which is in addition illuminated by a green LED. The LED is pulsed by an adjustable voltage and frequency. This setup is located in a light tight, temperature stabilized box. A fixed fraction of the LED light is fed, via a $3\,\mbox{m}$ long glass fiber with negligible temperature dependence, into the entrance window of the R669 photomultiplier which measures the \CsI\ crystals. A sketch of the setup is shown in Fig.~\ref{fig:refsetup}.
Variations in the LED light and in the gain of the photomultiplier of the reference system are corrected by the output of the light normal. In this way an error of better than $0.3\%$ for the total crystal light yield was achieved and verified by long-term comparisons of a few crystals.
At this level of accuracy, the temperature dependence of the crystal light yield becomes important. Therefore the crystals were located in a styrofoam box and temperature sensors were attached on the wrapping material. The light yield for each crystal was corrected by the measured temperature dependence of the light output (see section~\ref{sec:temp}).\\

\subsection{Photodiode Readout}
\label{sec:pd}

For the light readout by photodiodes, radioactive sources of higher photon energy are required. We have chosen \y\ with  $E_{\gamma}=898\keV$ and $1836\keV$. The scintillation light is read out by two PIN silicon photodiodes coupled to the crystal rear. Two different types of photodiodes from Hamamatsu were used, S~2744-08 with an active window size of $10\times 20\smm$ for direct readout and S~3588-03 (mod~5400) with an active area of $3\times 30\smm$ for wavelength shifter readout. These types have an active layer of $300\mum$ thickness and are SiN passivated. The diodes are operated with a reverse bias voltage of $-60\V$ with the cathode grounded. Typical values for the dark current of $3\nA$ and for the capacity of $85\pF$ ($45\pF$) are reported by the manufacturer for the types S~2744-08 and S~3588-03 (mod~5400), respectively. We measured the dark current of an initial sample of 110 diodes of type S~2744-08 in a temperature controlled environment using a Keithley pA-meter to be $(2.8\pm 0.5)\nA$. The same device was used for a measurement of the diode capacity. Thereto a rectangular generator was set up using the diode as a capacitor determining the frequency of the generator. This frequency was converted into a voltage which was measured as a current over a $5\MOhm$ resistor. The result was $(83.4\pm 0.7)\pF$ for 110 diodes.
The quantum efficiency of both types are 85\% (90\%) at $560\nm$ ($650\nm$) \cite{ref:hamamatsupd}.\\

The optical coupling of the diodes to the crystal was varied and will be described in section~\ref{sec:coupling}. The \py s of both photodiodes were measured individually by independent electronic readout chains~\cite{ref:bn216, ref:bn242}. In order to amplify the small pulses typically between 5000 and 8000 photoelectrons for a $\gamma$ source energy of $1.8\MeV$, charge sensitive preamplifiers, Canberra 2003T, were coupled via short co-axial cables to the diodes. Crystal, photodiodes, and preamplifiers were located in a light tight box which also provided an electromagnetic shield. From the preamp output each signal was fed into a spectroscopic shaping amplifier, Canberra AFT~2025, with $2\mus$ (Gaussian) shaping time and finally DC coupled into a peak sensitive ADC, Montedel Laben model 8215. The pile-up output of the amplifiers was connected to the corresponding ADC input. The ADC are read out with the same TMCA based DAQ system as used for the photomultiplier setup. The photopeak regions in both pulse-height spectra were fit with a combination of a Gaussian and an exponential in order to determine the peak position. \\

The ADC channel scale was calibrated in absolute electric charge by using the photodiodes as solid state detectors. The peak position for $59.5\keV$ photons from a \am\ source, directly located at the photodiode, was measured. Those photons, when absorbed in the silicon layer, create one electron-hole pair per $3.6\eV$ deposited photon energy. Therefore, the mean peak position of the photon line, which was determined by a fit of a Gaussian and a linear background in the peak region, corresponds to 16480 created photoelectrons. Thus the \py\ \PY\ of a crystal - photodiode combination, with dimensions $\emi/\MeV$, can be determined by the ratio of peak positions of scintillation light and photon conversion in the diode according to
\begin{equation}
\PY = \frac{\mbox{peak position}(\gamma\ \mbox{source})}{\mbox{peak position}(\am)}
                  \times \frac{16480\emi}{E_{\gamma}}.
\label{eq:py}
\end{equation}
\begin{figure}[t]
\begin{tabular}{cc}
\mbox{\epsfig{file=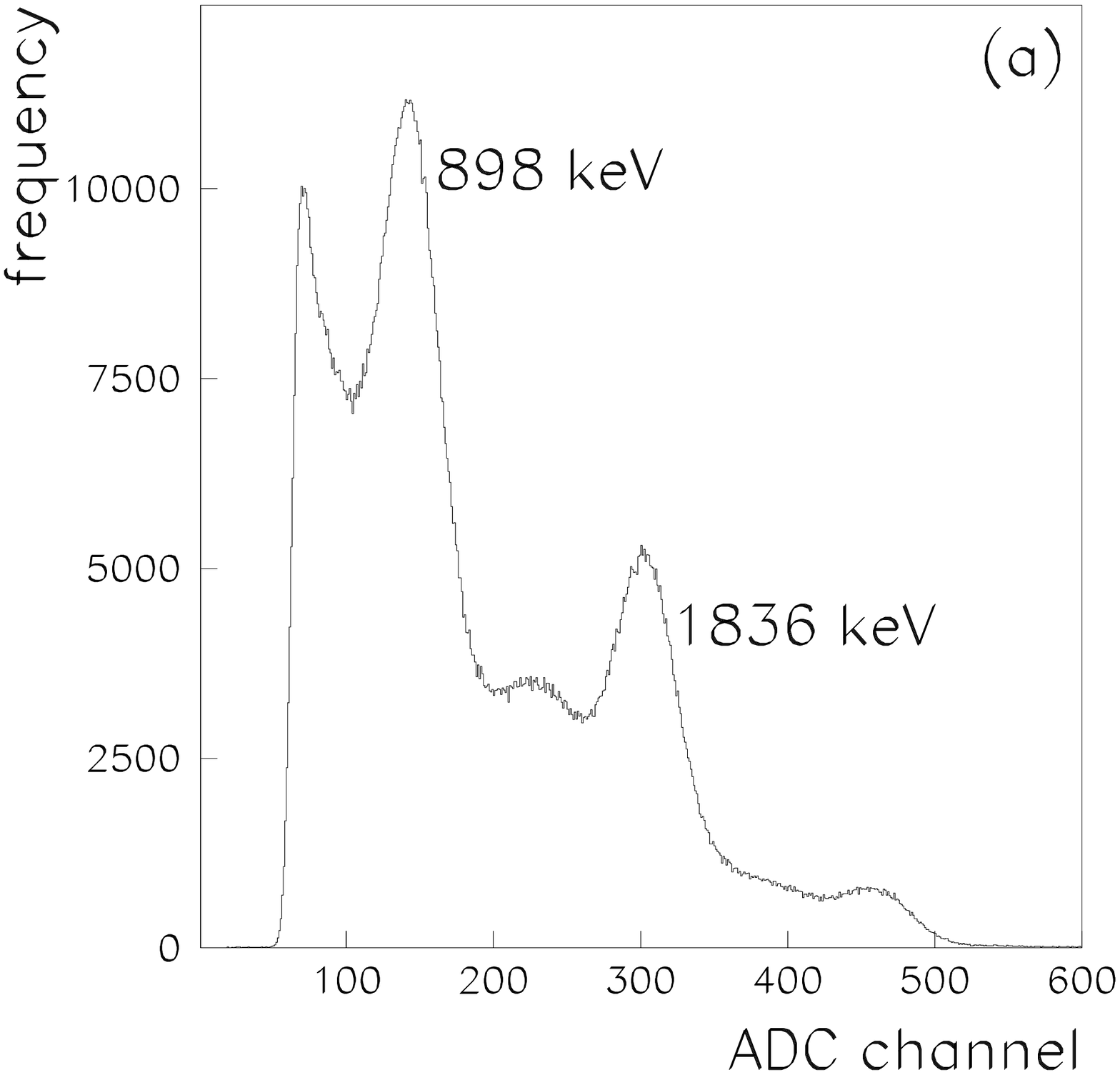,width=0.48\textwidth}} &
\mbox{\epsfig{file=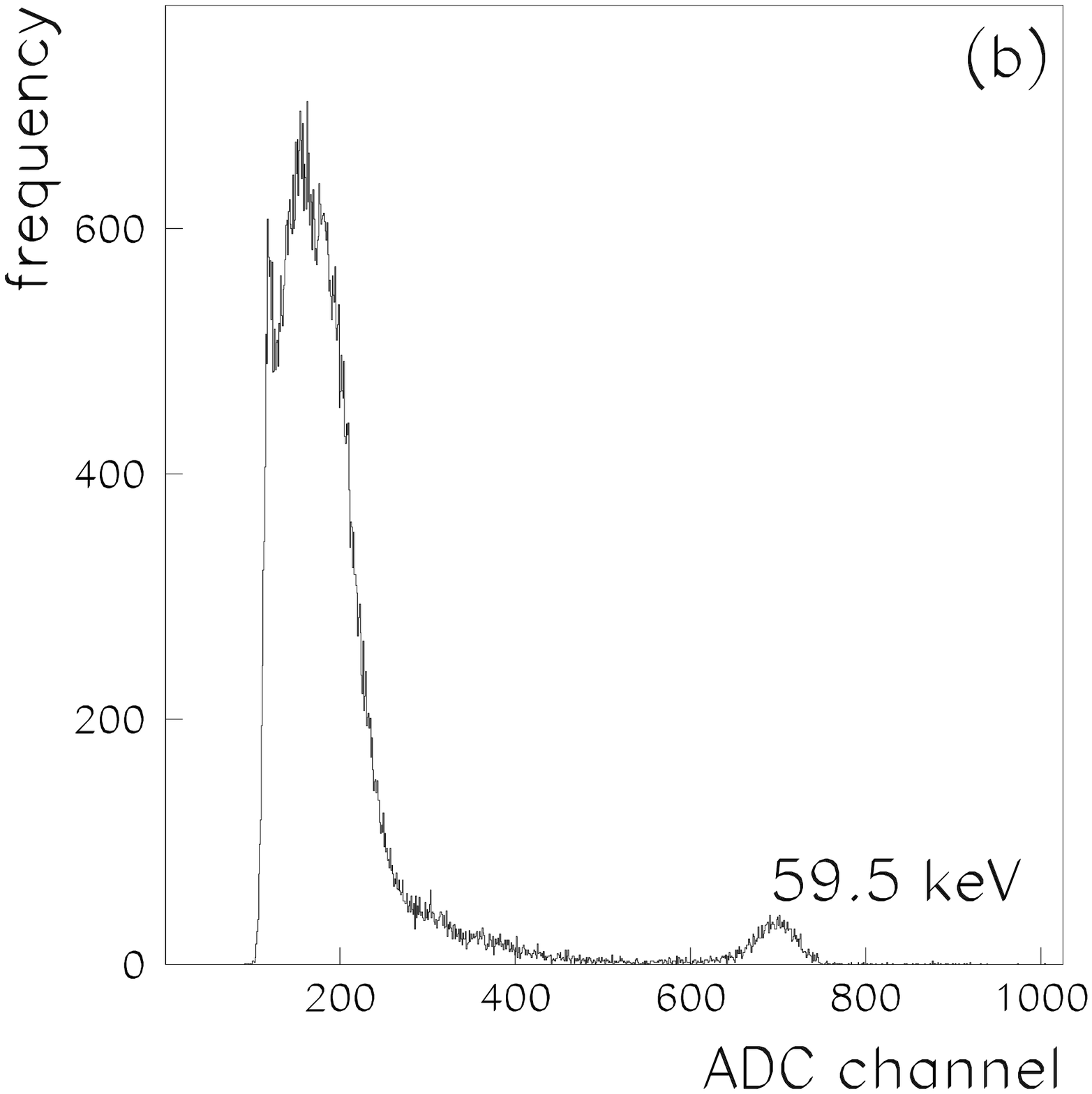,width=0.48\textwidth}} \\
\end{tabular}
\caption[]{CsI(Tl) scintillation spectrum as measured with one photodiode for a \y\ source (left) and spectrum of photoelectrons generated by \am\ $59.5\keV$ photons in the active material of the photodiode (right).}
\label{fig:pdspectra} 
\end{figure}
Typical spectra are shown in Fig.~\ref{fig:pdspectra}.
The \py s of both diodes mounted to one crystal were measured individually and  added later in order to quote the \py\ of the crystal - diodes setup. This method allowed the measurement of \py s in a wide range with an accuracy of $\pm 100\emi/\MeV$. The crystals used for this study had photoelectron yields between
 $5000$ and $9000\emi/\MeV$.\\

The ratio of the Gaussian peak width and peak position of the $59.5\keV$ \am\ line was used to determine the \enc\ \ENC :
\begin{equation}
\ENC = \frac{\mbox{peak width}(\am)}{\mbox{peak position}(\am)}
                  \times 16480\emi
\label{eq:enc}
\end{equation}
Typical RMS values for photodiode types S~2744-08 and S~3588-03 (mod~5400) and
the amplification / digitization chain as described above are $600\emi$ and 
$450\emi$, respectively. The \ene\ \ENE\ follows from Eqs. \ref{eq:py} and 
\ref{eq:enc},
\begin{equation}
\ENE = \frac{\ENC}{\PY} .
\label{eq:ene}
\end{equation}
Typical RMS values are 120 to $180\keV$ depending on crystal light yield.

\subsection{\sloppy Comparison of Photomultiplier and Photodiode Readout}
\label{sec:pmpd}

\begin{figure}[t]
\begin{tabular}{cc}
\mbox{\epsfig{file=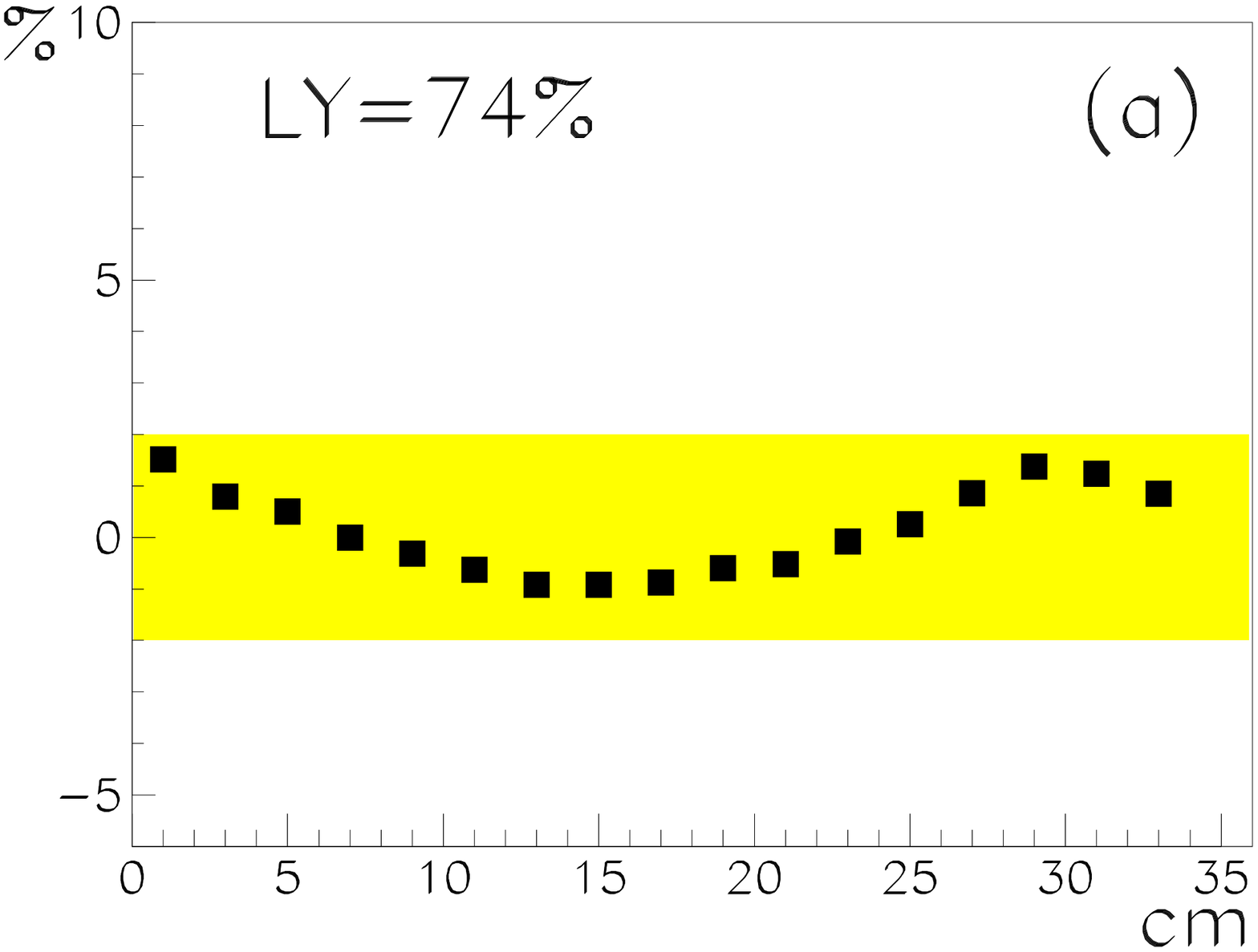,width=0.48\textwidth}} &
\mbox{\epsfig{file=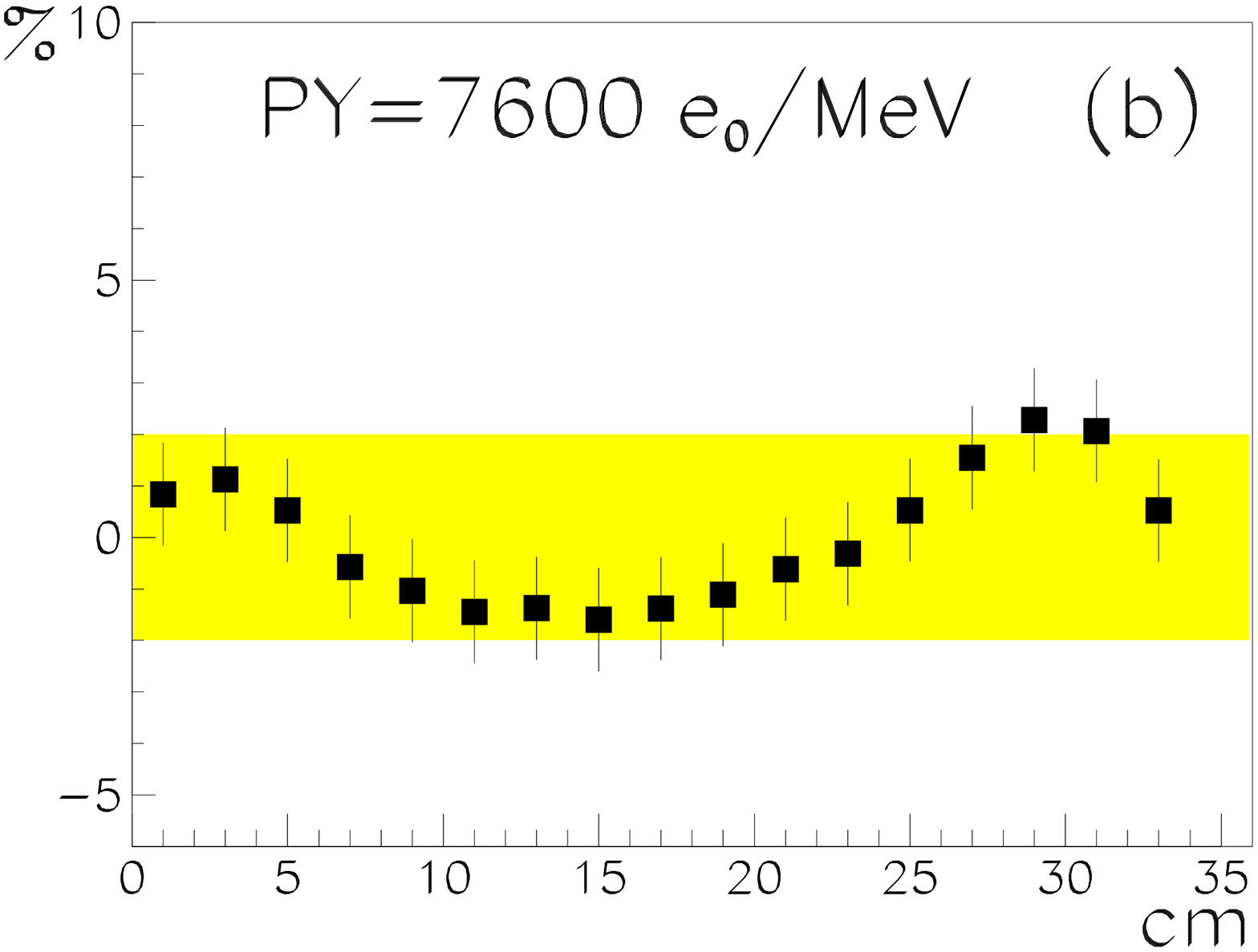,width=0.48\textwidth}} \\
\end{tabular}
\caption[]{Comparison of uniformity scans of a CsI(Tl) crystal with photomultiplier (a) and photodiode (b) readout using a \na\ source and a modified collimator (see text). The readout devices were mounted at crystal position $36\cm$. The plotted values are the relative deviation of light (photoelectron) yield at a given position from the average yield in percent. In case of the photomultiplier, the symbol height corresponds to the measurement error of the individual points. The shaded area indicates uniformity variations within $\pm 2\%$.}
\label{fig:scanpmpd} 
\end{figure}
In order to compare the uniformity of a crystal determined with different readout devices as well as to relate the \ly\ measured with the photomultiplier setup and the \py\ measured with photodiode readout, a uniformity scan with both setups was performed. Since the photon energy of \cs\ is too low to determine the photopeak position with photodiodes, the $1275\keV$ line of \na\ was used with a modified collimator consisting of two ${5\times 10\times 18\cm}^3$ lead blocks with a $5\mm$ wide gap. The source was placed in this gap $8\cm$ above the crystal surface. The crystal, wrapped in two layers of Tyvek 1056D and $5\mum$ aluminized mylar foil (see section~\ref{sec:wrapping}), was first scanned with the setup described in section~\ref{sec:pm} and then with the modified collimator and the \na\ source, which gave identical results within the measurement errors for light yield and uniformity. Finally, two photodiodes S~2744-08 were glued to a 1mm thick lucite plate, which in turn was glued to the center of the crystal rear face. The remaining rear area of the crystal was covered with a reflector, also consisting of Tyvek 1056D. The scan was repeated with the same stepwidth, but using the photodiode readout electronics. Fig.~\ref{fig:scanpmpd} shows a very similar uniformity behaviour and allows the determination of the corresponding \py\ for a given \ly . Here $\LY = 74\%$ corresponds to $\PY = 7600\emi/\MeV$. This relation was verified later with a larger number of crystals.\\

\begin{figure}[t]
\begin{tabular}{cc}
\mbox{\epsfig{file=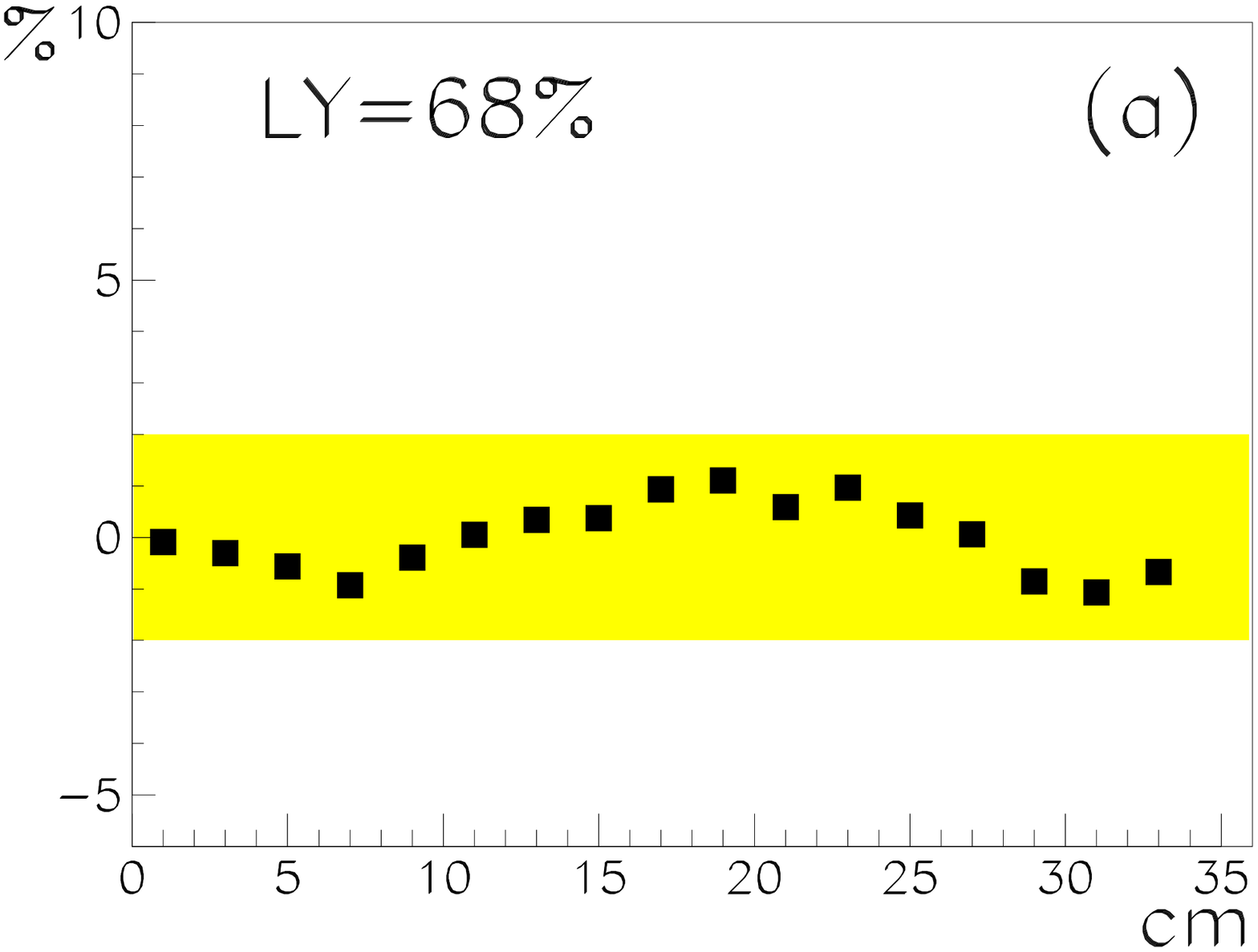,width=0.48\textwidth}} &
\mbox{\epsfig{file=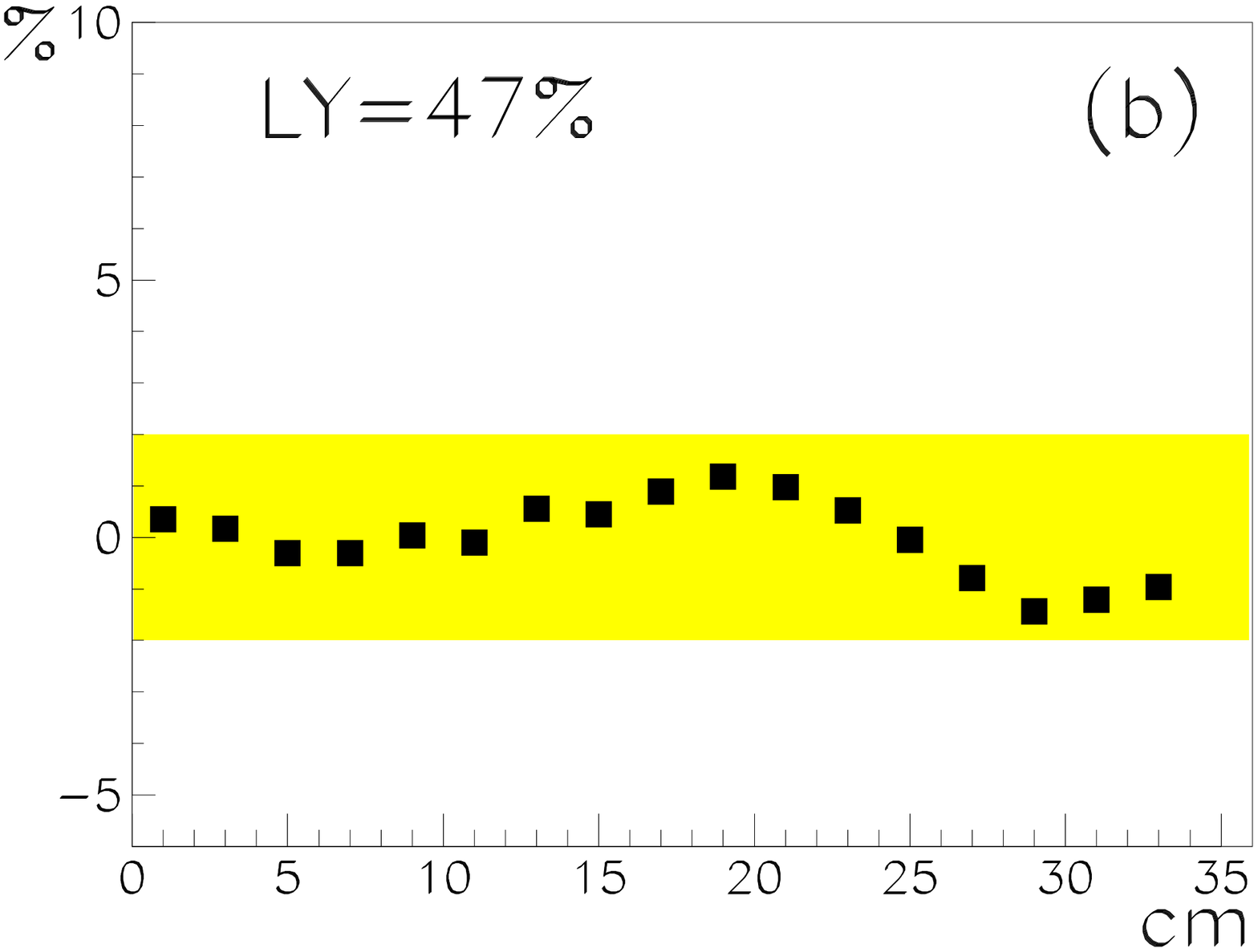,width=0.48\textwidth}} \\
\end{tabular}
\caption[]{Photomultiplier scan of a crystal using the whole crystal rear surface (a) and covering all but an area of $30\times 30\smm$ as in case of the photodiode readout (b).}
\label{fig:rearrefl} 
\end{figure}

The uniformity of the crystal light output does not depend on the area covered by the readout device, as shown in Fig.~\ref{fig:rearrefl}.
All other photodiode measurements described in the remainder of this article were performed following the prescription of section~\ref{sec:pd} without uniformity scan, whereas all photomultiplier uniformity scans were measured with the monitoring setups as described in section~\ref{sec:pm}. The uniformity plots will always have the photomultiplier located at position $36\cm$ (the larger crystal area). The plotted values are the relative variation of light yield at a given position from the average yield in percent.

%%%%%%%%%%%%%%%%%%%%%%%%%%%%%%%%%%%%%%%%%%%%%%%%%%%%%%%%%%%%%%%%%%%%%%%%%%%%%

\section{Optimization of Crystal Light Yield and Uniformity}
\label{sec:optim}

In this section the influence of type, quality, and thickness of wrapping material on crystal light yield and uniformity is described. A method to improve the uniformity of the light output is presented. These studies were performed with the photomultiplier setup which gives more precise results. The influence of optical properties of different coupling materials between crystal and photodiodes is described in section~\ref{sec:coupling}.

\subsection{Crystal Wrapping}
\label{sec:wrapping}

\begin{figure}[t]
\begin{tabular}{ccc}
\mbox{\epsfig{file=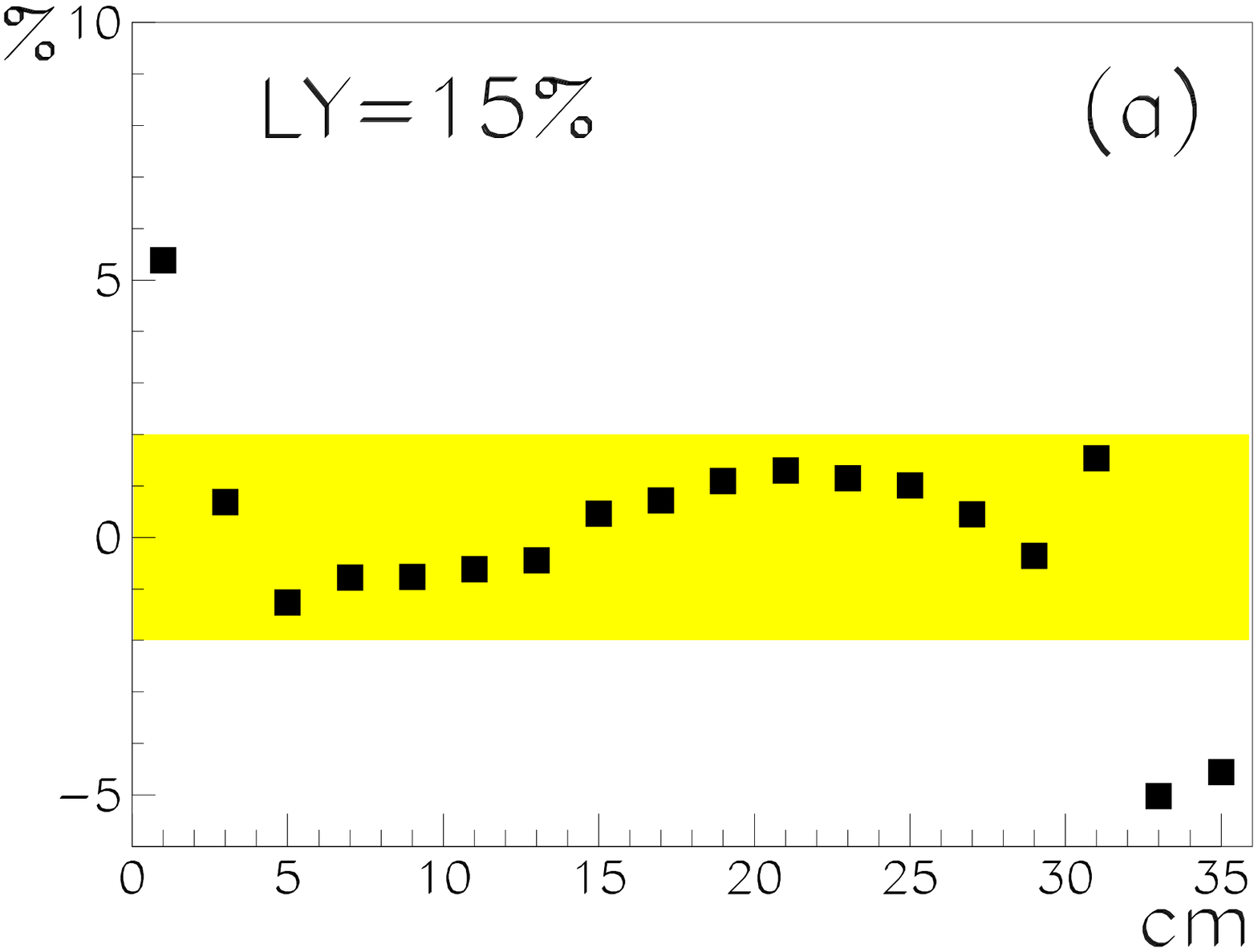,width=0.31\textwidth}} &
\mbox{\epsfig{file=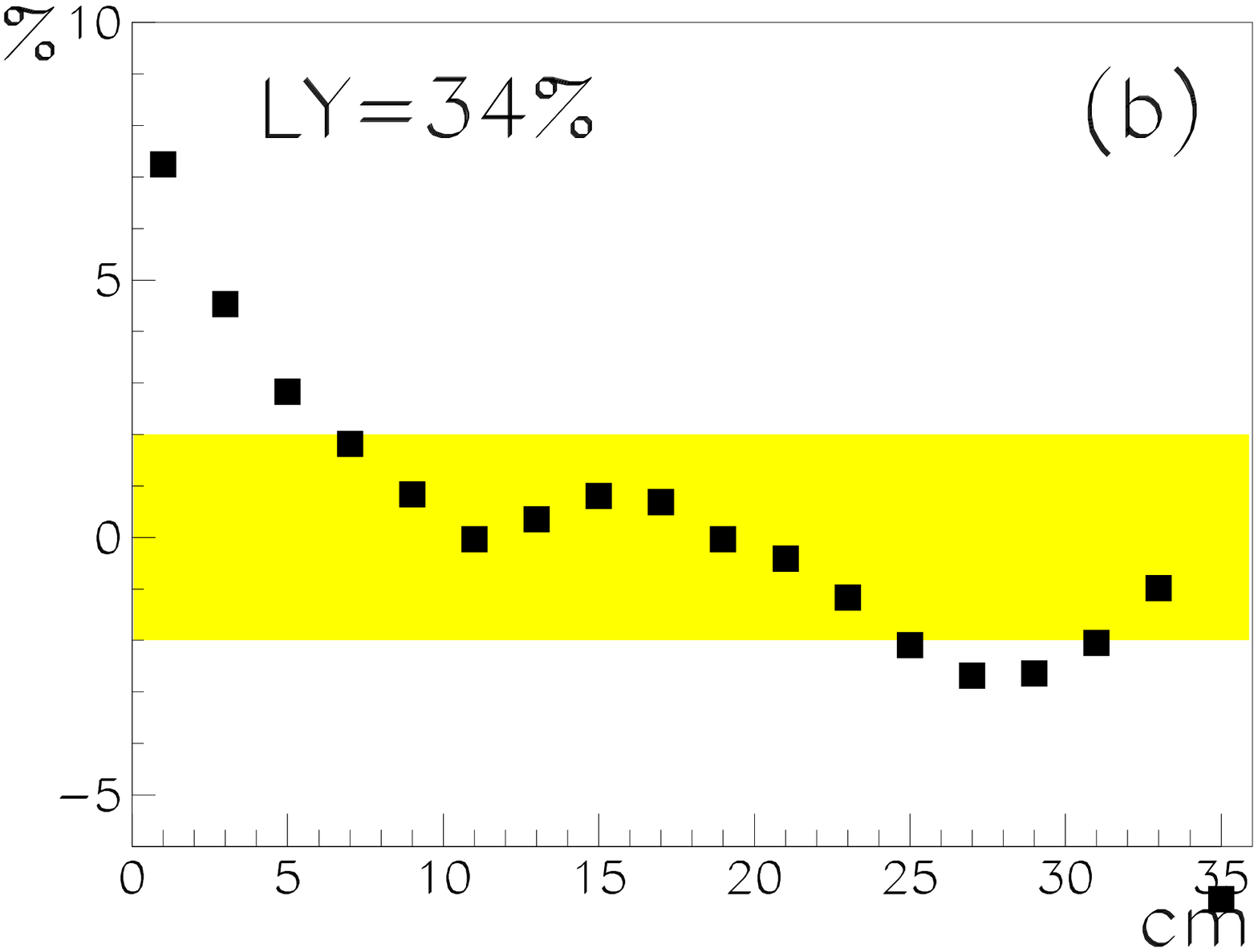,width=0.31\textwidth}} &
\mbox{\epsfig{file=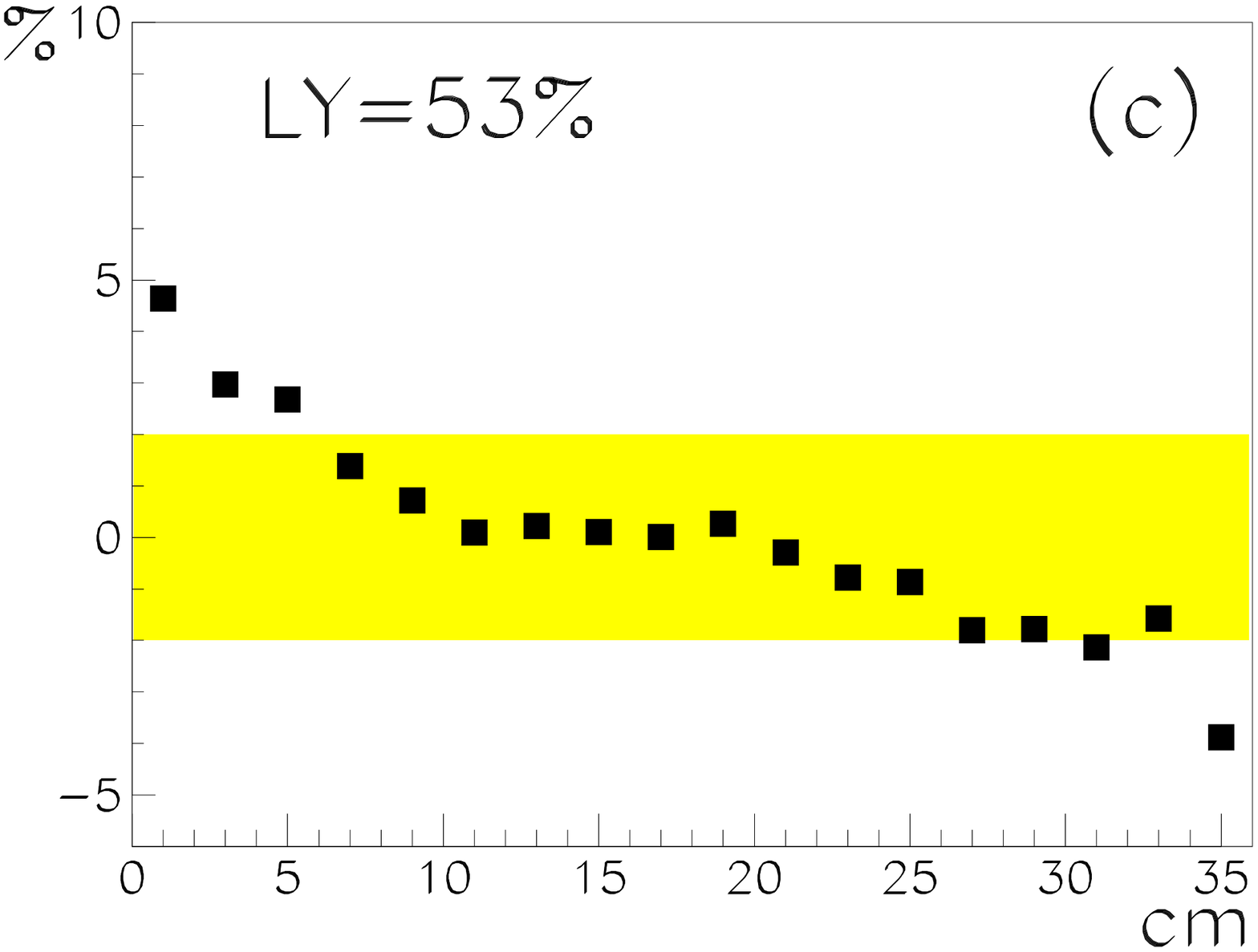,width=0.31\textwidth}} \\
\mbox{\epsfig{file=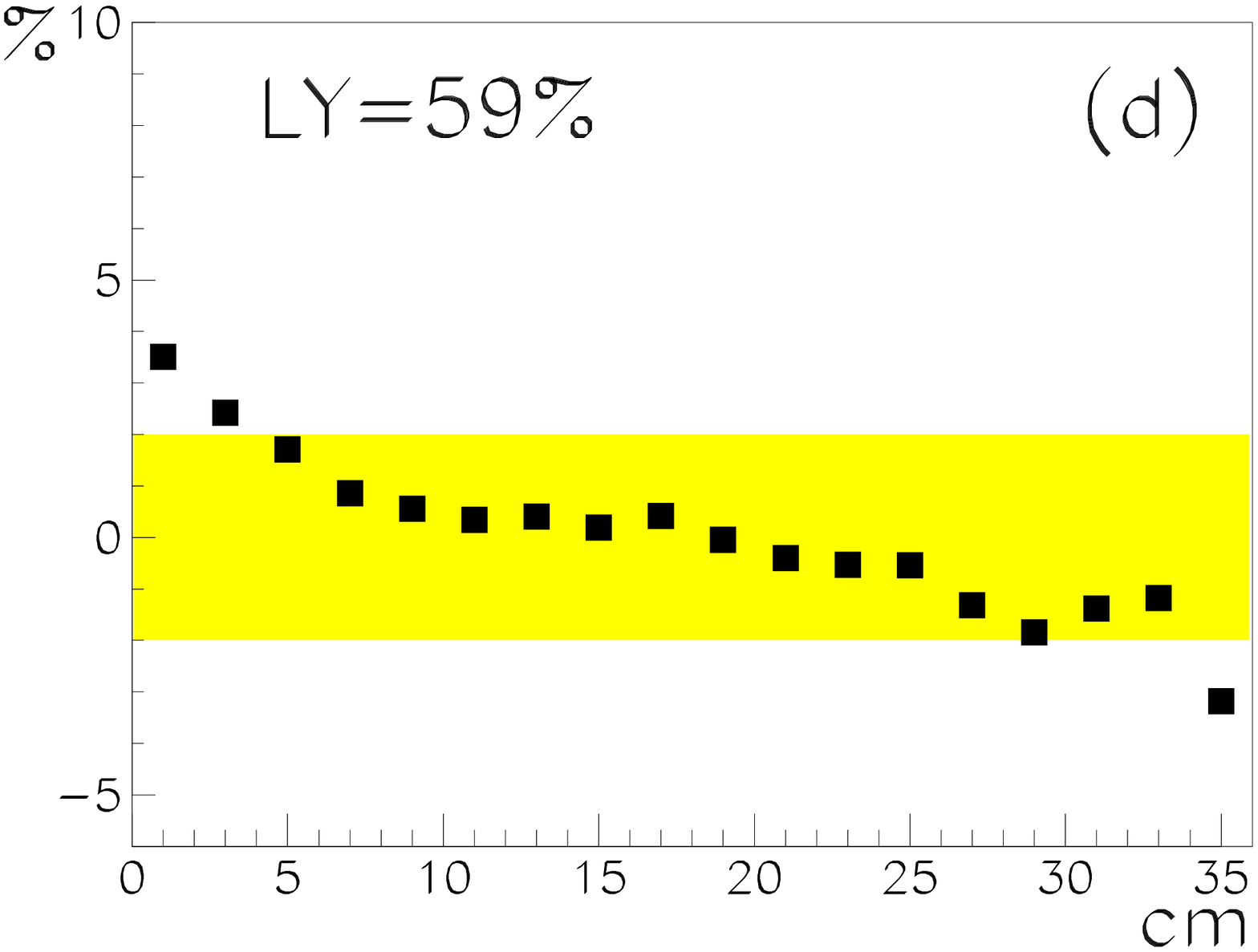,width=0.31\textwidth}} &
\mbox{\epsfig{file=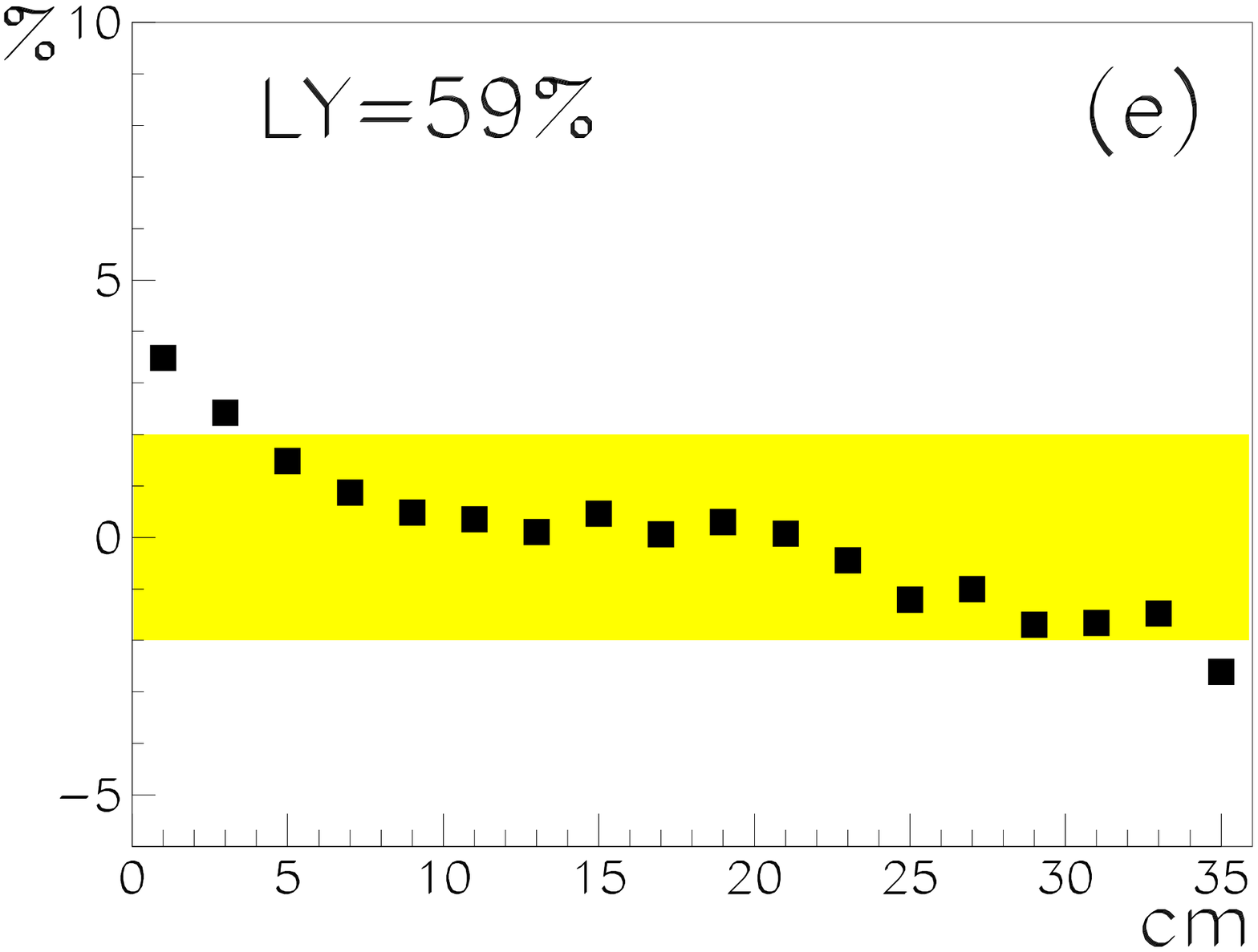,width=0.31\textwidth}} &
\mbox{\epsfig{file=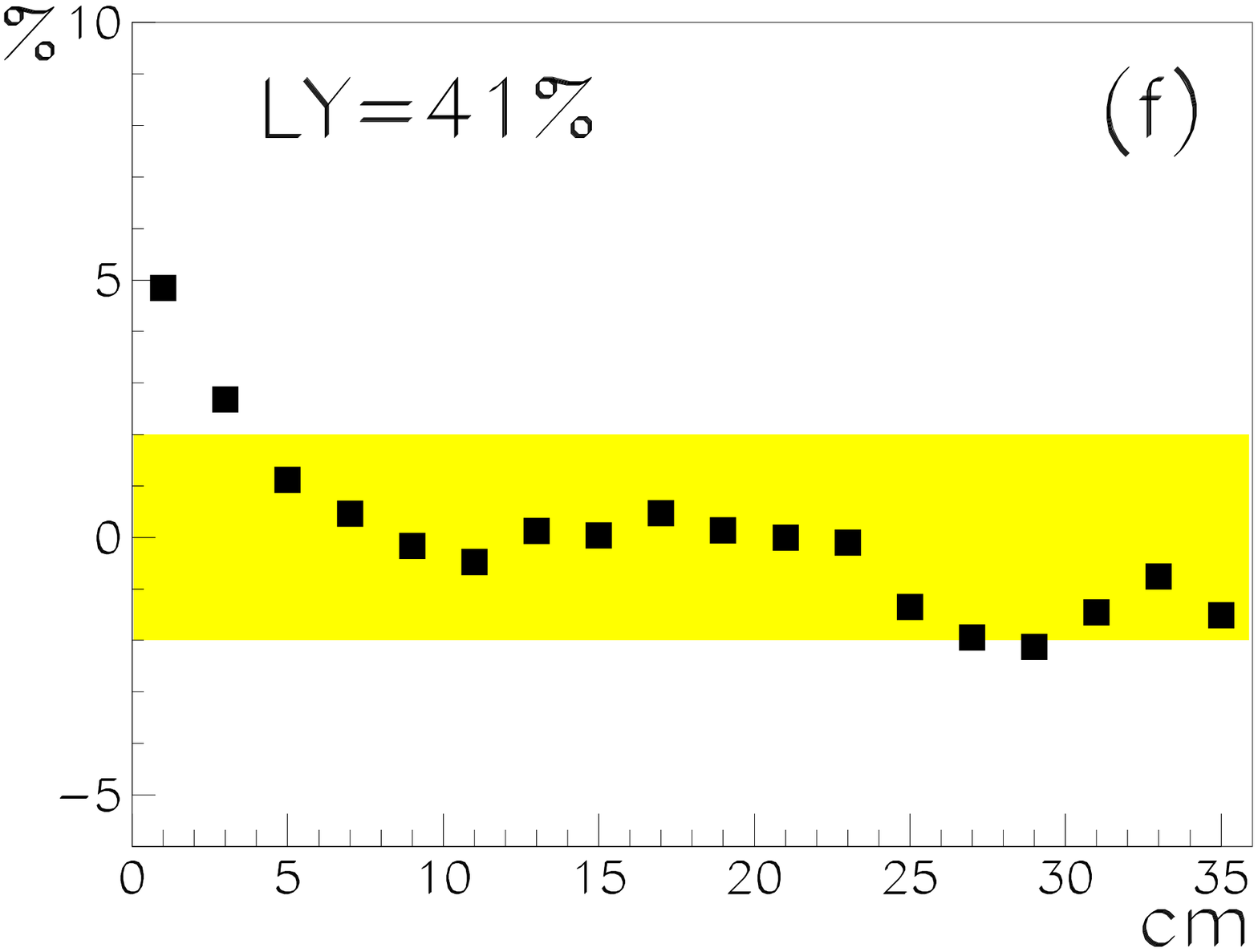,width=0.31\textwidth}} \\
\end{tabular}
\caption[]{Wrapping studies with Teflon PTFE. No wrapping (a), one layer 
of $38\mum$ Teflon (b), three layers of $38\mum$ Teflon (c), four layers of 
$38\mum$ Teflon (d), three layers of $38\mum$ Teflon and $5\mum$ aluminum foil (e), and one layer of $200\mum$ Teflon (f).}
\label{fig:teflon} 
\end{figure}

After measuring the light output achieved by only internal reflection of scintillation light in a crystal with surfaces polished by the manufacturer, see Fig.~\ref{fig:teflon}(a), consecutive layers of $38\mum$ thick Teflon PTFE are added. With more material wrapped around the crystal, the \ly\ increases and uniformity improves. Using an ideal reflecting $5\mum$ thick aluminum foil after three layers of Teflon results in the same total \ly\ as use of an additional layer of Teflon (Fig.~\ref{fig:teflon}(d) and (e)) . The comparison of four layers Teflon ($\sim 160\mum$ in total) with one layer of $200\mum$ thick Teflon shows that the multilayer configuration gives a higher light yield, Fig.~\ref{fig:teflon}(d) and (f).
The gain in light yield and uniformity by adding additional layers of material
is compromised by the increasing amount of dead material between crystals.\\

\begin{figure}[t]
\begin{tabular}{cc}
\mbox{\epsfig{file=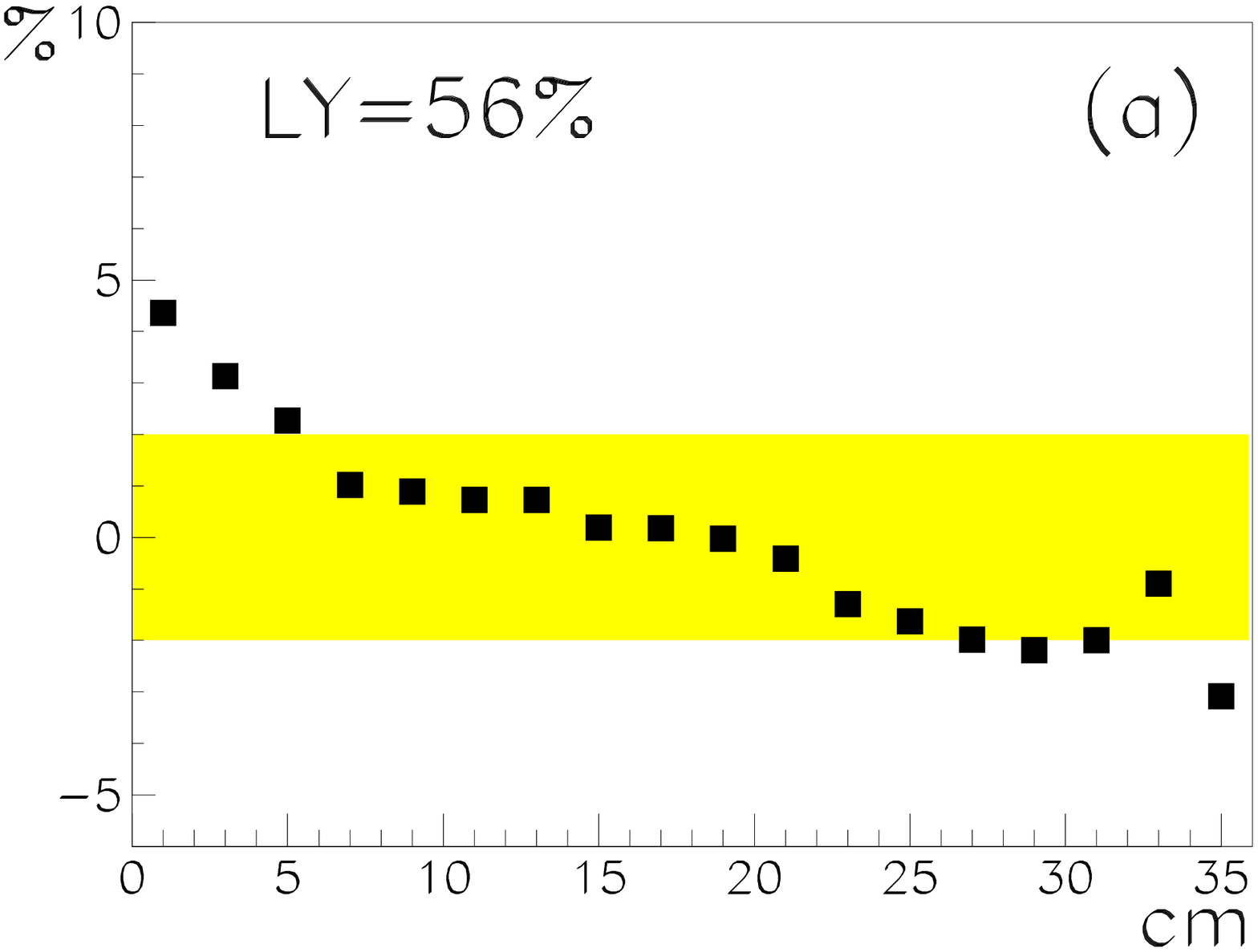,width=0.48\textwidth}} &
\mbox{\epsfig{file=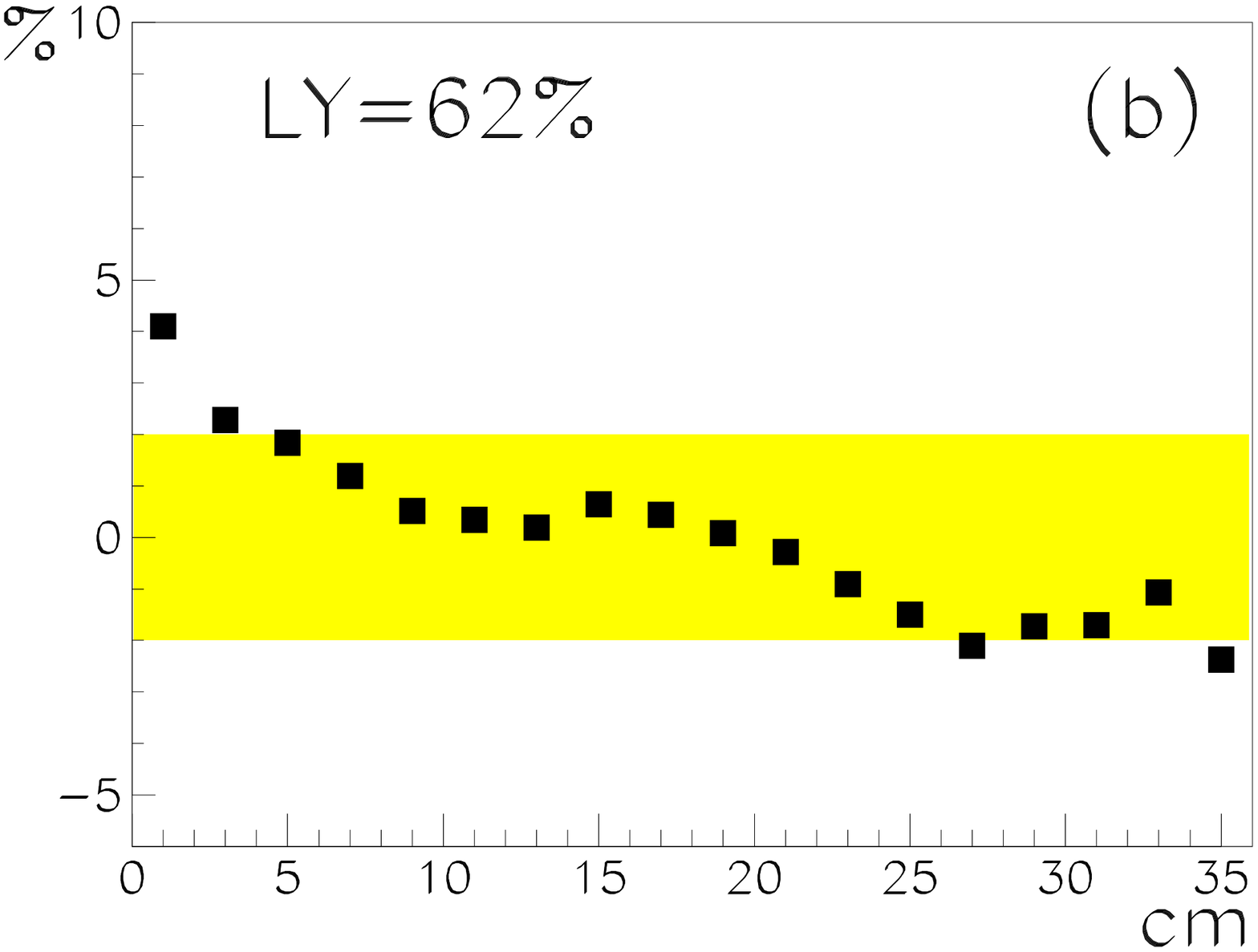,width=0.48\textwidth}} \\
\mbox{\epsfig{file=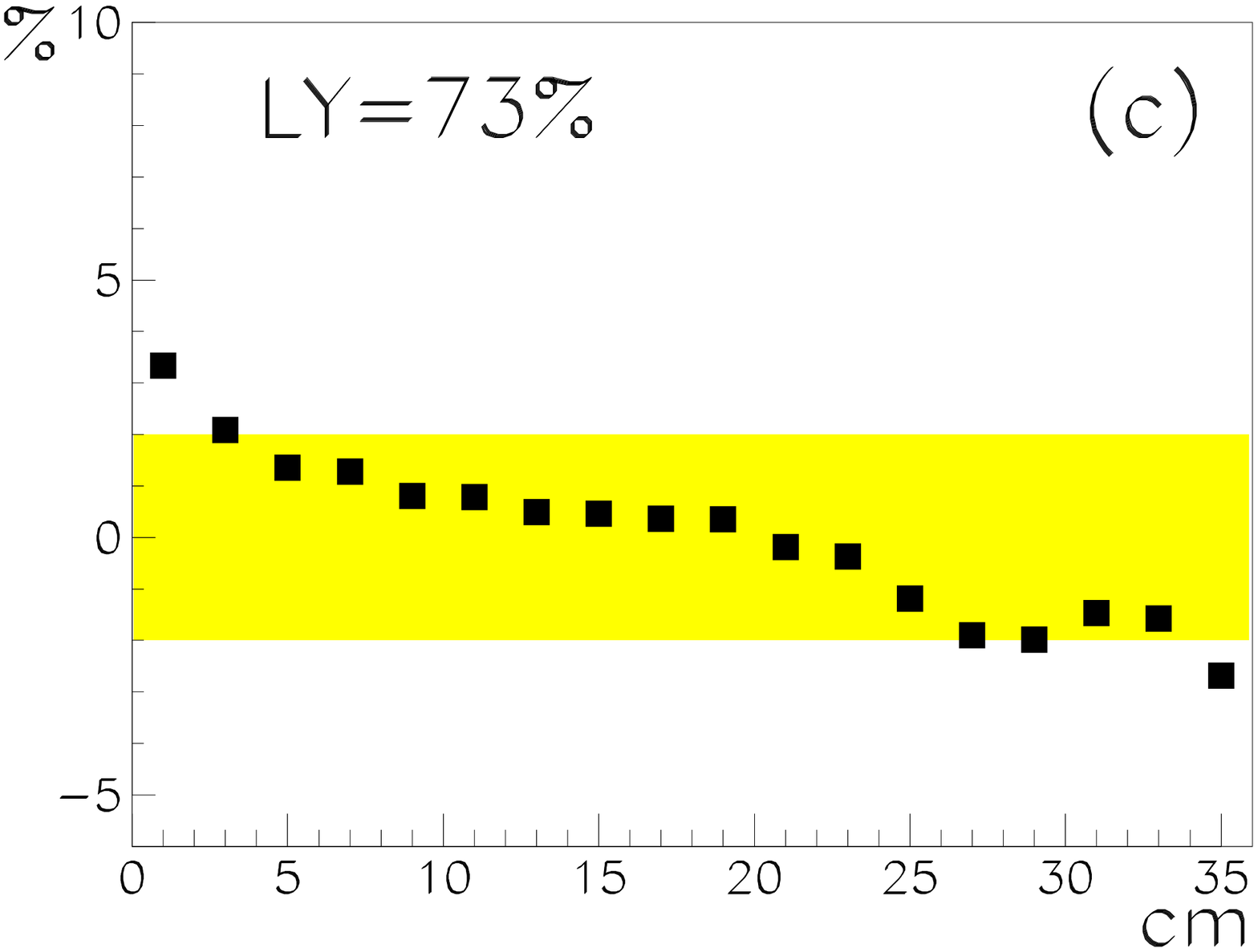,width=0.48\textwidth}} &
\mbox{\epsfig{file=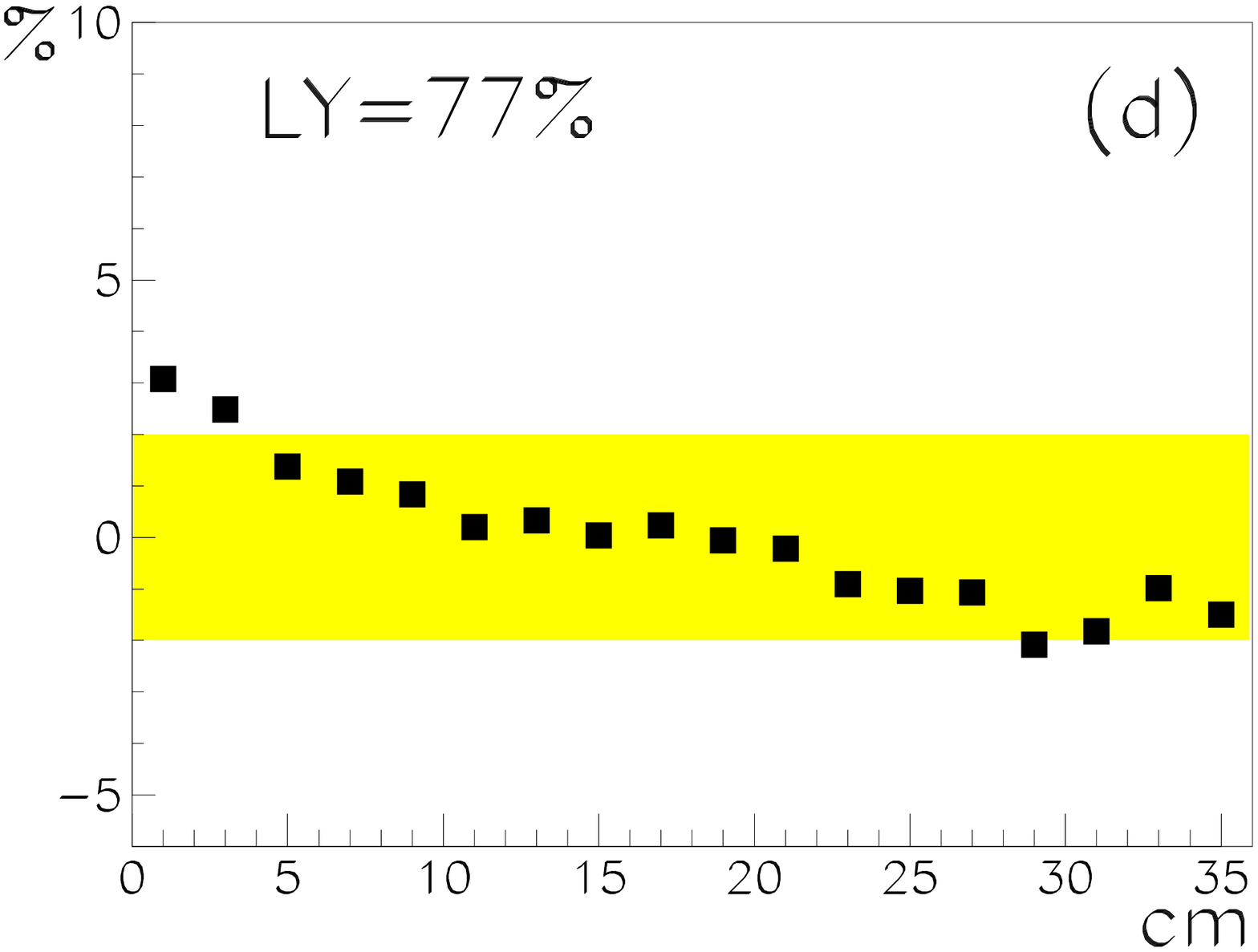,width=0.48\textwidth}} \\
\end{tabular}
\caption[]{Wrapping studies with Tyvek 1056~D for the same crystal as in Fig.~\ref{fig:teflon}. One and two layers without, (a) and (b), and with,(c) and (d), additional $5\mum$ aluminum foil.}
\label{fig:tyvek} 
\end{figure}

Since Teflon PTFE shows an unfavourable longtime behaviour (multilayers tend to
be pressed together, thus forming a monolayer with reduced reflecting 
properties) and is difficult to handle because of its adhesive properties, 
DuPont Tyvek was investigated as an alternative~\cite{ref:bn206, ref:bn241}. 
This material is a porous, chalk-loaded polyethylene fleece, which is supplied
in different qualities and thicknesses.
For the Tyvek types 1025D, 1056D, and 1059D total \ly\ values were measured, 
which are comparable to the results for multilayers of Teflon of similar total 
thickness. Fig.~\ref{fig:tyvek} shows uniformity plots for one and two layers 
of Tyvek 1056~D, $160\mum$ per layer, with and without an additional aluminum
foil layer.
With increasing material thickness, the uniformity again improves. The aluminum foil not only gives higher light yield values, but also provides electromagnetic shielding of the crystal.

\subsection{Uniformity Tuning}
\label{sec:tuning}

\begin{figure}[t]
\begin{tabular}{cc}
\mbox{\epsfig{file=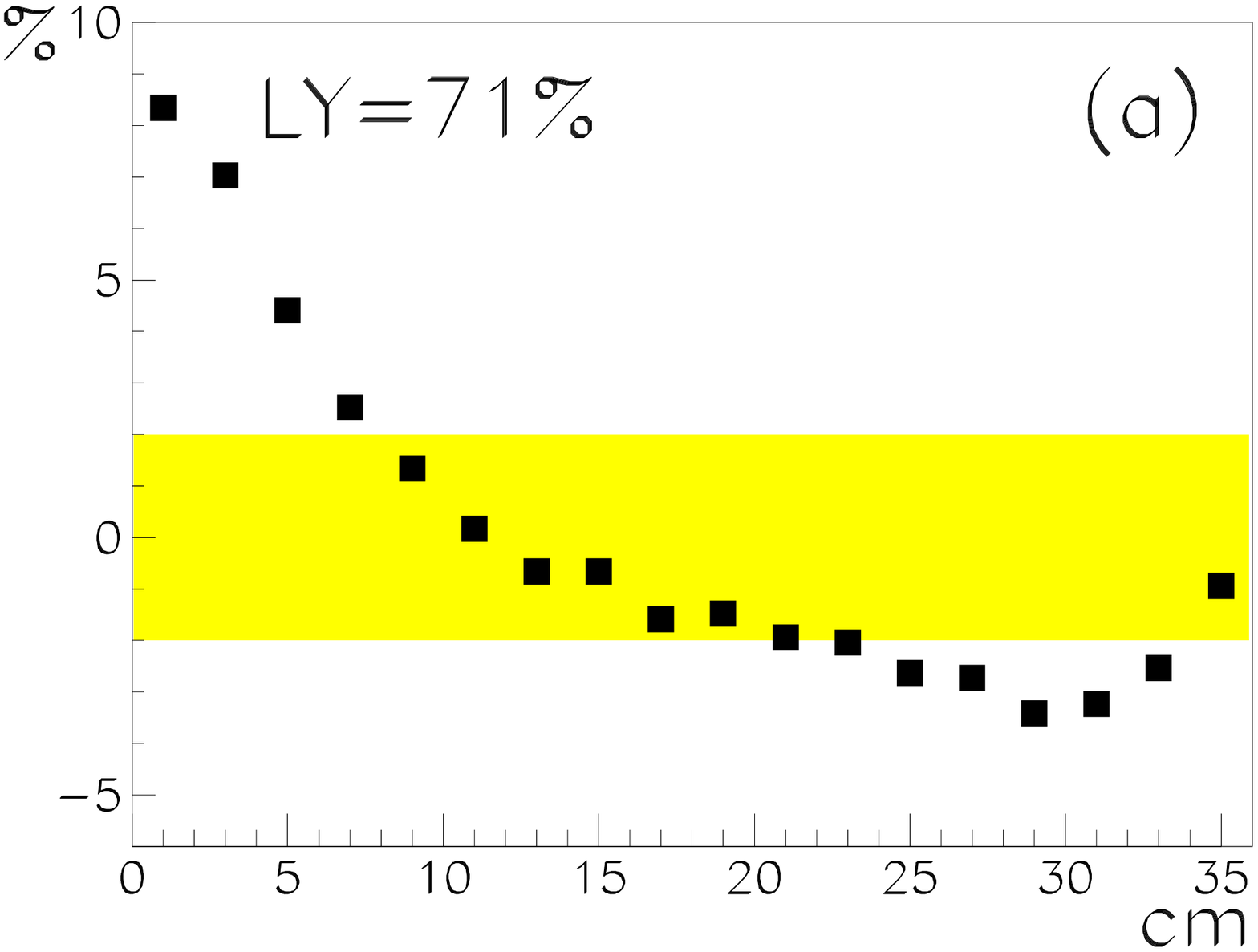,width=0.48\textwidth}} &
\mbox{\epsfig{file=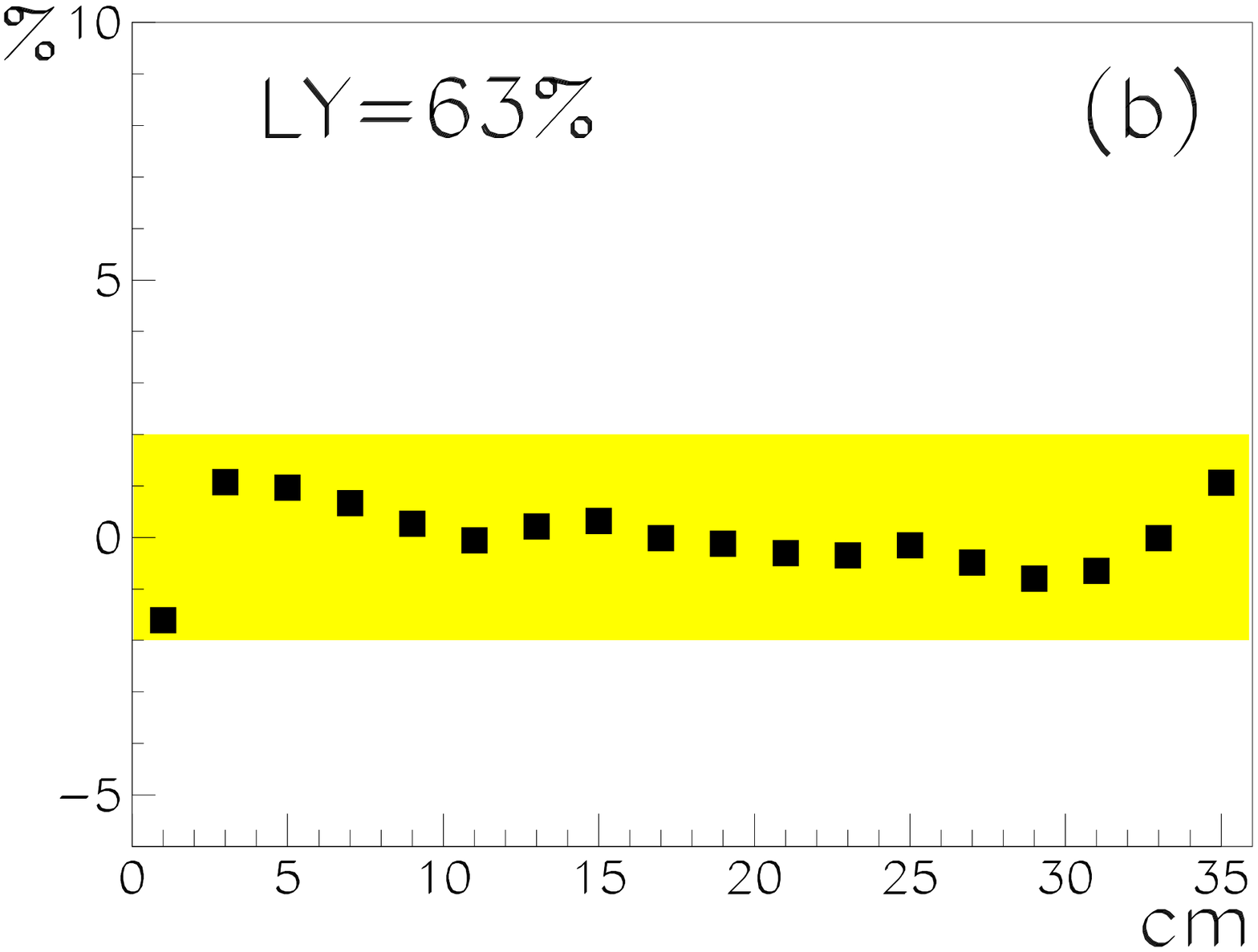,width=0.48\textwidth}} \\
\end{tabular}
\caption[]{Uniformity plot for a crystal before (a) and after (b) application of a black strip at the wrapping material at position $0\ldots 7\mm$. The crystal total \ly\ decreases using this method.}
\label{fig:tune} 
\end{figure}

A uniform crystal response is important for the linearity of a detector and the energy resolution, since the position of the shower maximum varies with the incident energy and fluctuates for fixed energies.
An additional advantage of Tyvek is the possibility to improve the uniformity
of the light output of crystals by changing local absorption properties at the crystal surface without machining the crystal itself. This is done by blackening the Tyvek at appropriate positions. An example is shown in Fig.~\ref{fig:tune}.

\subsection{Optical Coupling}
\label{sec:coupling}

Two general types of coupling of the photodiodes to the crystal rear surface were investigated: wavelength shifter and direct readout~\cite{ref:bn216, ref:bn242}. 

\subsubsection{Wavelength Shifter Readout}
\label{sec:wls}

The wavelength shifters consisted of fluorescent dyes, dissolved in $3\mm$ thick lucite (PMMA) carrier plates covering the whole crystal rear face. Both investigated types (BASF Lumogen F red 300 and Roehm 2466 P) show similar optical properties (several absorption maxima at about $440\nm$, $530\ldots 540\nm$, and
$560\ldots 570\nm$; fluorescence emission at $\sim 625\nm$), thus effectively collecting the crystal scintillation light and transforming it into a wavelength range where the quantum efficiency of the photodiode is higher (see section \ref{sec:pd}). They had the full size of the crystal rear surface and were coupled via a small ($\sim 0.5\mm$) airgap to the crystal. The diodes were glued onto two adjacent narrow sides of the wavelength shifter. The rear area was covered by a diffuse reflecting material (Tyvek), whereas all the lateral faces not equipped with photodiodes were painted with polyurethane based reflective paint NE~561.
The adhesive used to glue the photodiodes S~3588-03 (mod~5400) to the wavelength shifter was Cargille Meltmount with an index of refraction of 1.53, well matched to the corresponding values for PMMA and the entrance window of the photodiodes (both $\sim 1.5$).\\ 

Using two photodiodes on adjacent edges of the wavelength shifter did not double the \py\ compared to the readout with one photodiode, but increased it by a factor of 1.5 only. The second photodiode does not transfer all absorbed light into measureable charge, whereas a reflecting material at position of the second photodiode redirects a part of the light to the first photodiode.  This is compared in Table~\ref{tab:wls} with an absorber ($30\mm$ black Scotch tape) placed at the position of the second photodiode. In this case, the light yield is only two thirds of the value achieved with one photodiode and a reflector. 
\begin{table}[t]
\begin{center}
\begin{tabular}{|lccc|}
\hline
& 2 photodiodes & 1 photodiode & 1 photodiode \\
&               & + reflector  & + absorber \\
\hline 
\PY\  $[\emi/\MeV]$  & 5900 & 3900 & 2600 \\
\ENE\ $[\keV]$       &  110 &  115 &  175 \\
\hline
\end{tabular}
\caption[]{Photoelectron yield and \ene\ (RMS) for a $6\times 6\scm$ 
wavelength shifter with one and two attached photodiodes.}
\label{tab:wls}
\end{center}
\end{table}
The maximum \py\ of $6050\emi/\MeV$ for two photodiodes was achieved with an 
octagonal wavelenght shifter with $30\mm$ on an edge (= length of photodiode 
window) and a diameter of $60\mm$ between parallel sides. This corresponds to 
an \ene\ of $105\keV$ for the crystal.

\subsubsection{Direct Readout}
\label{sec:direct}

In case of the direct readout the main focus was the search for the best 
coupling medium between the crystal and the photodiode entrance medium 
with respect to \py\ and reliability.
In order to diminish  long-time effects such as migration of ions from the 
crystal to the diodes active material, which could result in a decrease of 
the \py , additional transparent coupling plates were 
investigated~\cite{ref:bn242, ref:bn236}. Their optical properties are given 
in Table~\ref{tab:plate}.
\begin{table}[t]
\begin{center}
\begin{tabular}{|cc|}
\hline
Coupling & Index of   \\
Plate    & Refraction \\
\hline
Bor-Silicate Glass & 1.50 \\
Lucite (PMMA) & 1.50 \\
Lexan (PC) & 1.59 \\
Polystyrene & 1.60 \\
\hline
\end{tabular}
\caption[]{Optical properties of coupling plates for direct readout. The 
transmission for all types is $>95\%$ at wavelengths above $450\nm$.}
\label{tab:plate}
\end{center}
\end{table}
 Because of the difference in the indices of refraction of the crystal 
($n_{CsI}=1.79$) and the entrance window of the photodiode ($n_{PD}=1.50$) 
materials with higher refractive index are favoured~\cite{ref:bn270}. This is 
also the reason for typically $\sim 15\%$ lower \py\ in the case of a small 
airgap between crystal and photodiodes.
Table~\ref{tab:glues} shows the various optical adhesives which have been 
investigated, and typical ranges of \py s for a class of adhesives.
Comparing the results in Table~\ref{tab:glues}, optical grease and epoxies give
the highest \py s. Because of its poor longterm behaviour (it tends to flow and
to change color), optical grease cannot be considered for CsI(Tl) crystal based calorimeters.\\

The highest \py s were achieved with the last three epoxies of Table~\ref{tab:glues}. They all have a refractive index of 1.56. Using BC~600 for both glue joints (coupling plate to crystal and photodiodes to coupling plate) the three coupling plate types Lucite, B-Si-Glass, and Polystyrene give very similar \py s ($\sim 8000\emi/\MeV$), whereas Lexan is worse ($7000\emi/\MeV$). Bor-Silicate Glass is very fragile and hard to machine and handle. EPOTEK 301-2 tended towards disconnecting from the crystal over longer time periods ($\sim $ months).  Using the combination photodiodes - BC~600 - lucite plate and testing the three epoxies for the coupling plate - crystal joint, EPILOX A17-01 is slightly favoured ($8150\emi/\MeV$ vs. $7900\emi/\MeV$). The best result was achieved when using a polystyrene coupling plate and EPILOX for both glue joints ($8450\emi/\MeV$). This corresponds to a \ene\ of $100\keV$, in the same range as in the case of the wavelength shifter readout. The lower \py\ of the latter is compensated by the lower intrinsic noise of the smaller diodes, see Eq.~\ref{eq:ene}. 
For all measurements the rear area of the crystal surrounding the photodiodes was covered by a diffuse reflecting plate (either Tyvek or reflective paint NE~561). Without these rear reflectors the measured \py s were about $30\%$ lower. 

\begin{table}[ht]
\begin{center}
\begin{tabular}{|lc|}
\hline
optical adhesive & $\PY$ $[\emi/\MeV]$\\
\hline
\underline{Grease} & 7000 to 8000 \\
BICRON BC 630 & \\
Rhodosil Silicones Pate  B 431 & \\
\hline
\underline{Silicone Pads} & 5000 to 6000 \\
General Electrics GE RTV 615 & $n\sim 1.45$ \\
BICRON BC 634 & \\
\hline
\underline{Silicone Glues} & 6000\ to 6800 \\
General Electrics RTV 108  &  $n\sim 1.45$\\
General Electrics RTV 118 &  \\
N\"unchritzer RTV Cenusil &  \\
\hline
\underline{Acrylic Glues} & 6500 to 7000 \\
Powa Bond 102 &  $n\sim 1.50$ \\
DELO Photobond 4455 &  \\
Loctite 349 &  \\
Degussa Agovit 1900 &  \\
Roehm Acryfix 200/117 &  \\
Forbo Helmitin 21003 &  \\
Vitralit 5638 &  \\
\hline
\underline{Epoxies} & 7000 to 8500 \\
Ciba Geigy XW 396/397 & $n\sim 1.55$ \\
Ciba Geigy AY 951/103 &  \\
Kodak HE 80 &  \\
Lens Bond F65 &  \\
Structalit 701 &  \\
EPOTEK 301-2 &  \\
BICRON BC 600 &  \\
EPILOX A17-01 &  \\
\hline
\end{tabular}
\caption[]{Optical adhesives investigated: The typical range of \py s was 
measured with two photodiodes S~2744-08 glued with BICRON BC~600 to a $1\mm$ 
thick lucite plate, which in turn was coupled with the corresponding adhesive 
to the crystal.}
\label{tab:glues}
\end{center}
\end{table}

%%%%%%%%%%%%%%%%%%%%%%%%%%%%%%%%%%%%%%%%%%%%%%%%%%%%%%%%%%%%%%%%%%%%%%%%%%%%%

\section{Temperature Dependence of Crystal Light Yield and Uniformity}
\label{sec:temp}

Different values for the temperature dependence of the \CsI\ \ly\ are 
reported in the literature. In most studies 
\cite{ref:valentine,ref:kobayashi} the light yield dependence was 
measured for small crystals in a wide temperature range 
($-100\degc$ to $+50\degc$) in coarse steps.
These authors report an almost constant crystal light yield between 
$10\degc$ and $30\degc$.
Here, we concentrate on a comparison of the crystal \ly\ 
for photodiode and photomultiplier readout near room temperature, where 
large calorimeters are operated~\cite{ref:bn355}.\\

Photomultipliers have their own own temperature dependence, the used type Hamamatsu R669 shows a temperature coefficient of $-0.3\%/\K$ at $550\nm$ \cite{ref:hamamatsupm}. However, its influence is corrected by the reference system described in section~\ref{sec:pm}. The crystal temperature was varied between $10\degc$ and $30\degc$. For these measurements, an airgap of $1\mm$ between photocathode and crystal was chosen in order to allow for thermal expansion of the crystal. The photomultiplier was operated at constant temperature, in this way further reducing the measurement errors. The procedure was repeated three times; after the last measurement crystal and photomultiplier were slowly heated to $30\degc$.\\

\begin{figure}[t]
\begin{tabular}{cc}
\mbox{\epsfig{file=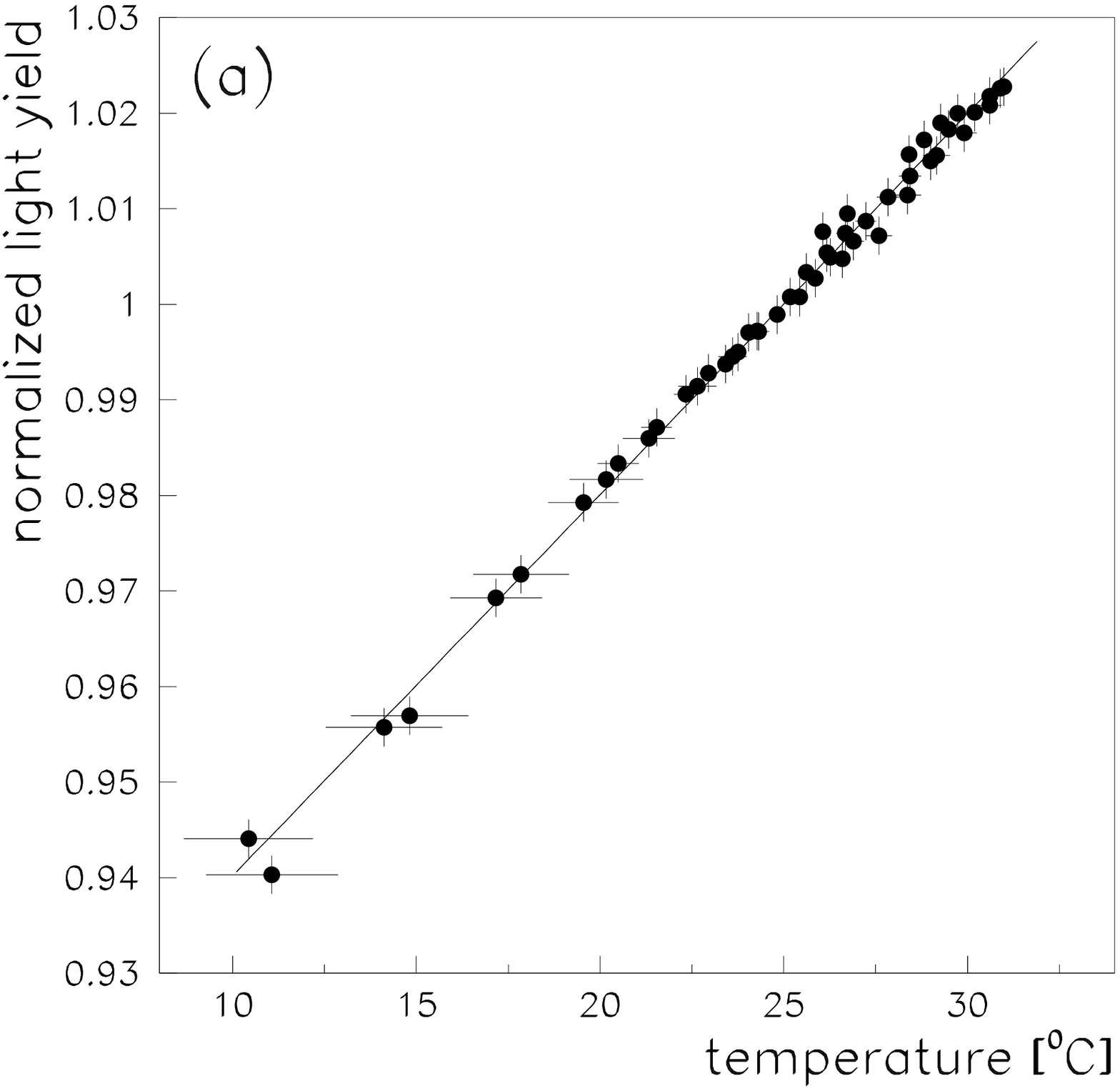,width=0.48\textwidth}} &
\mbox{\epsfig{file=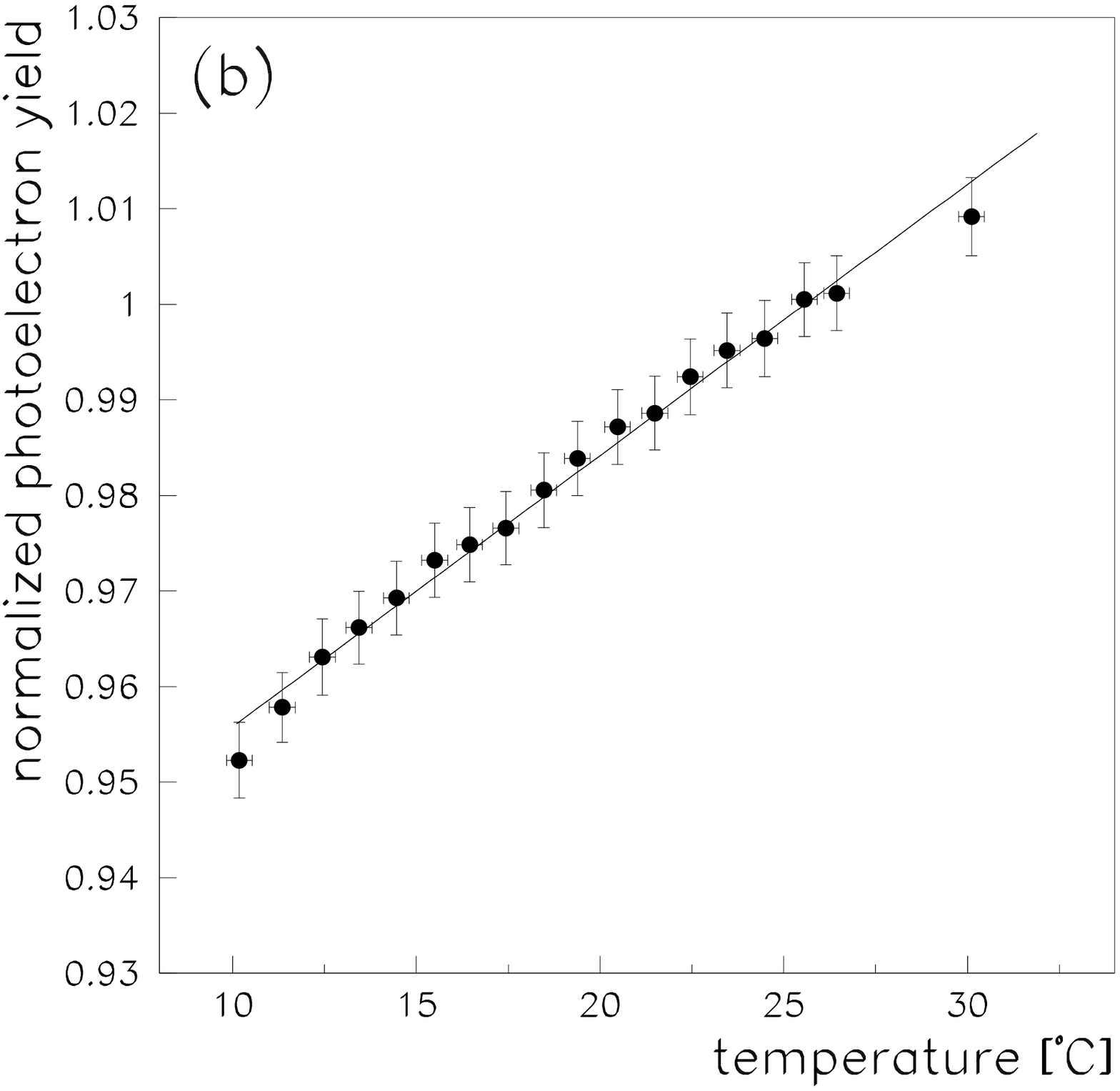,width=0.48\textwidth}} \\
\end{tabular}
\caption[]{Temperature dependence of crystal light yield for photomultiplier (a) and photodiode (b) readout. The temperature error bars indicate the temperature change during one measurement.}
\label{fig:temp} 
\end{figure}

In order to get a stabilized coupling of the photodiodes to the crystal, they were glued to the crystal using BICRON BC~600 epoxy and a $1\mm$ thick lucite plate and cooled down together with the crystal. Since the temperature dependence of the photosensitivity of the diodes is only $0.01\%/\K$ at $550\nm$~\cite{ref:hamamatsupd}, this will not distort the measured crystal properties. Fig.~\ref{fig:temp} shows the dependence of crystal light yield, normalized to $25\degc$. The crystal temperature coefficients were determined by a straight line fit to the data. The results are $+(0.40\pm 0.01)\%/\K$  for photomultiplier and $+(0.28\pm 0.02)\%/\K$ for photodiode readout, respectively. A reason for the difference between photomultiplier and photodiode readout could be the different spectral sensitivity of both systems, which indicates a decreasing temperature dependence of the scintillation light with longer wavelengths. This would also explain the higher value of $+0.6\%/\K$ found in~\cite{ref:grassmann}, where a blue-sensitive photomultiplier was used. For photodiode readout these authors find $+0.3\%/\K$ at $20\degc$
and report a flat maximum at $30\degc$~to~$40\degc$. A hint of the latter
can be seen in Fig.~\ref{fig:temp}(b).
The uniformity of crystal response did not change in the investigated temperature range.

%%%%%%%%%%%%%%%%%%%%%%%%%%%%%%%%%%%%%%%%%%%%%%%%%%%%%%%%%%%%%%%%%%%%%%%%%%%%%

\section{Influence of Radiation Damage on Crystal Light Yield and Uniformity}
\label{sec:rad}

The influence of radiation damage on the crystal light yield varies strongly
with intrinsic crystal properties (e.g. Thallium concentration, impurities,
absorption length). Therefore, the measured
crystal described below does not represent general \CsI\ properties; it merely
shows what kind of measurements and precision can be achieved with the 
setups described in section~\ref{sec:setup}~\cite{ref:bn355}.\\

The front face of a $36\cm$ long \CsI\ crystal was irradiated several times with increasing dose, using a point like $\sim 200\,\mbox{GBq}$ \co\ source, mounted at a fixed distance of $23\cm$. The time between individual irradiations was chosen to be one week, in order to allow for studies of relaxation processes.\\

The deposited energy dose in the crystal was determined by two independent 
dose measurements. At a low dose of about $70\rad$, four  LiF 
dosimeters with a thickness of $0.85\mm$ were put on the corners of the 
crystal front face. The measured dose rate for LiF was 
$(307 \pm 16)\rad/\mbox{h}$. All following measurements were scaled by the 
irradiation time according to this measurement.
In order to check the dose extrapolation at the highest individual dose of 
about $7\krad$, four Alanine dosimeters with a thickness of $0.987\mm$
were put at the same position and a dose rate of $(282 \pm 8)\rad/\mbox{h}$ 
was measured. Owing to the smaller error of the last measurement all other 
were scaled by the irradiation time according to this measurement.
In order to get the energy dose deposited at the crystal front face by 
photons of $1.17\MeV$ and $1.33\MeV$, the 
energy doses quoted in this section were evaluated for CsI using
the ratio of the mass energy absorption coefficients for photons of $1.25\MeV$
for CsI \mbox{($\mu_{en}/\rho=2.402 \cdot 10^{-2}$~cm$^2$/g)} 
and for Alanine \mbox{($\mu_{en}/\rho=2.878 \cdot 10^{-2}$~cm$^2$/g)} \cite{ref:dose}.
During the measurements, radiation doses from about $10\rad$ up to $7\krad$ in 
steps of about a factor of two were deposited at the front face of the crystal.\\

\begin{figure}[t]
\begin{tabular}{cc}
\mbox{\epsfig{file=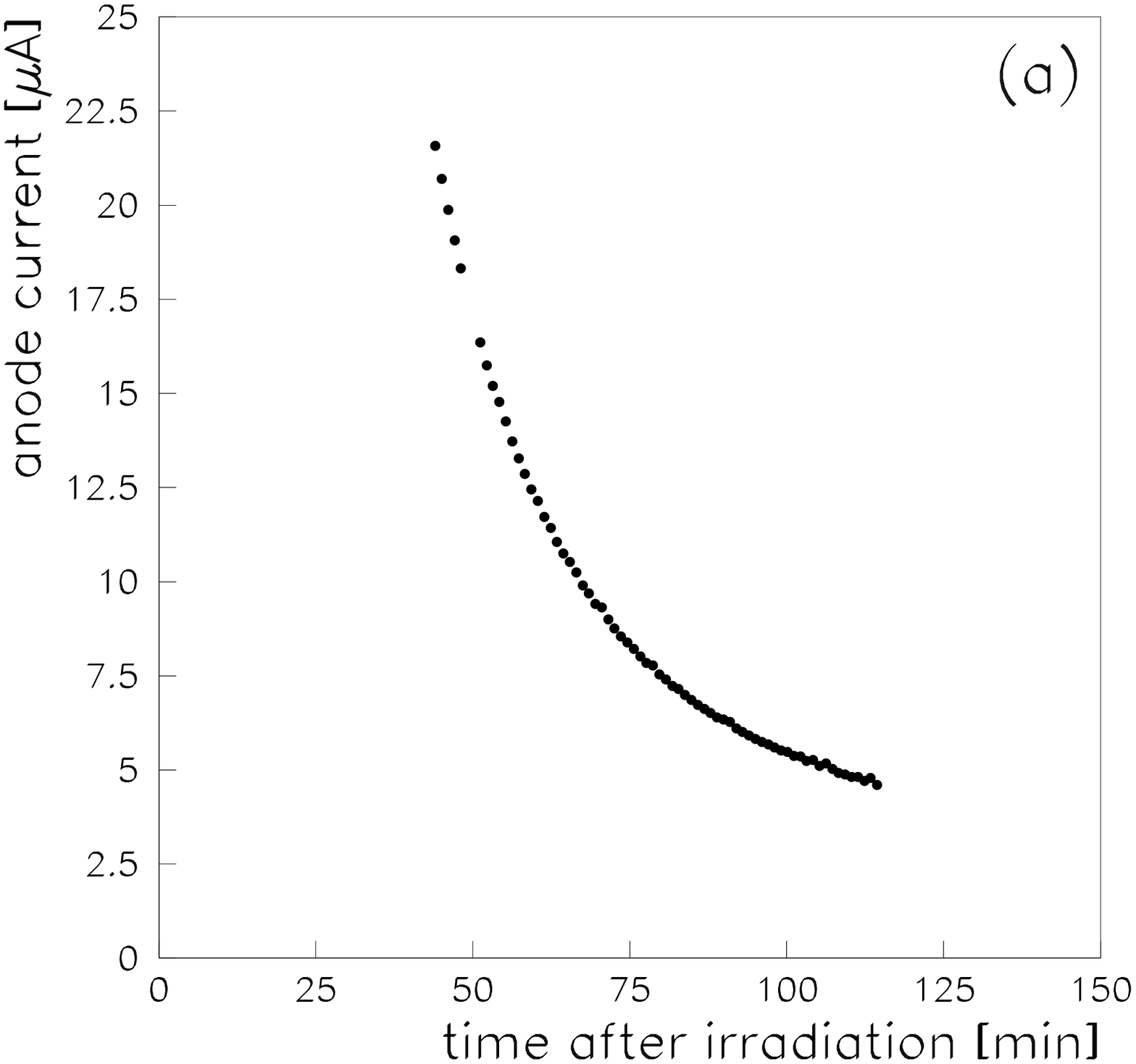,width=0.48\textwidth}} &
\mbox{\epsfig{file=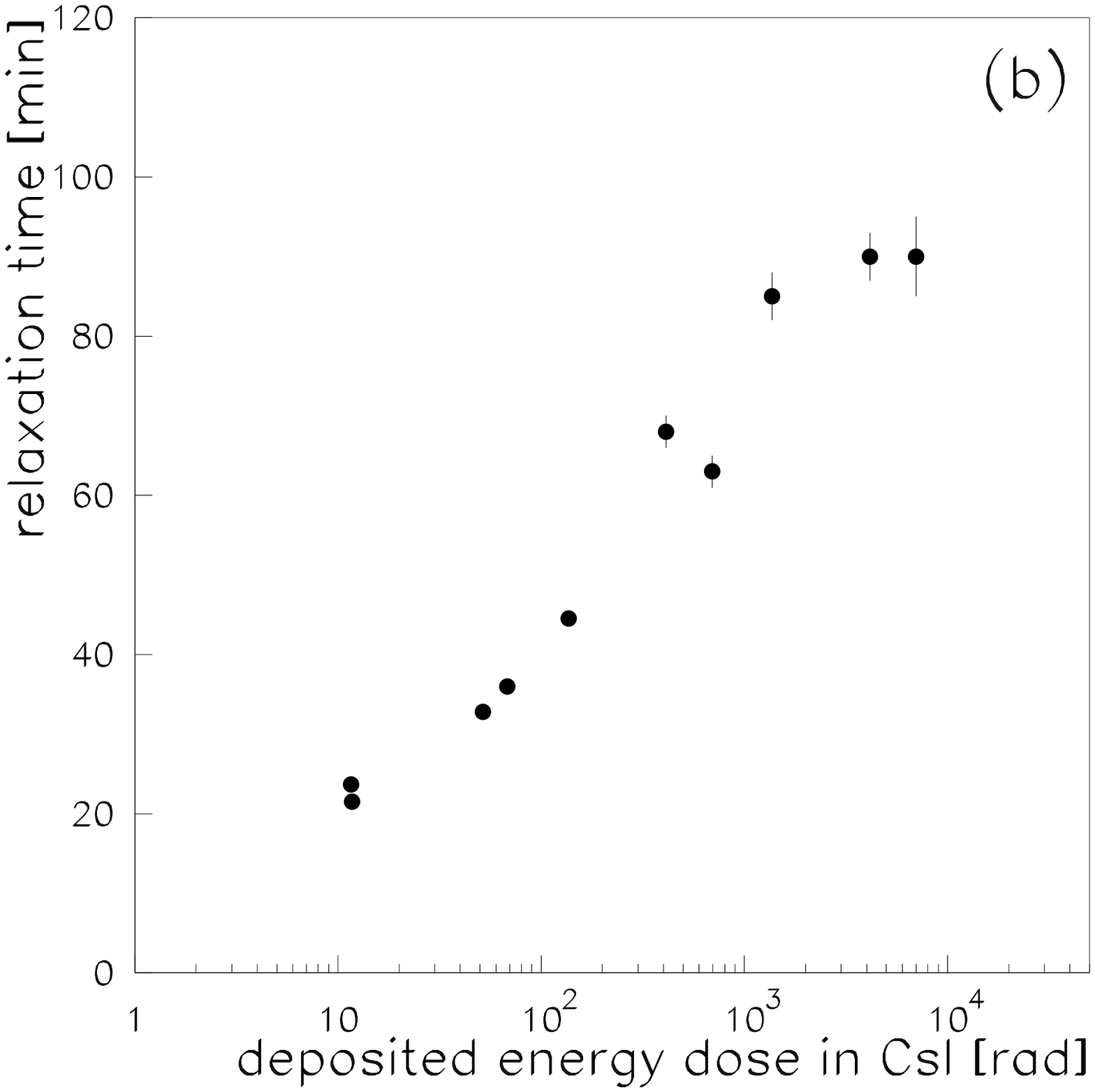,width=0.48\textwidth}} \\
\end{tabular}
\caption[]{Time dependence of phosphorescence light after irradiation at $12\rad$ (a) and relaxation time constants from fits of an exponential and a constant term to the time dependence measurements after all steps (b).}
\label{fig:phos} 
\end{figure}

After each irradiation step, the intensity of phosphorescence light 
(afterglow) was measured as the anode current of the photomultiplier
which was DC-coupled to 
a voltage/current device. To protect the photocathode, these measurements were
started after the phosphorescence light level corresponded to less than 
$40\muA$ anode current at a voltage of $1\kV$. Fig.~\ref{fig:phos}a shows
the current measurements after an irradiation of $12\rad$. 
Fig.~\ref{fig:phos}b shows the dependence of the phoshorescence time constant 
$\tau$ on the deposited energy dose, defined by $I(t)=I_0+I_1\exp{(-t/\tau)}$. 
The relaxation times $\tau$ rise from $20\m$ to $90\m$ for doses 
increasing from $12\rad$ to $7\krad$.
The constant currents $I_0$ in the fit of the time dependence increase from
 $\sim 3\muA$ to $\sim 22\muA$ over this dose range. Repeated measurements 
six days after each individual irradiation show values between $100\nA$ and 
$400\nA$, whereas the value before any irradiation was about $80\nA$.
These measurements indicate additional relaxation time constants for radiation 
induced phosphorescence on the order of days.\\

\begin{figure}[t]
\begin{tabular}{cc}
\mbox{\epsfig{file=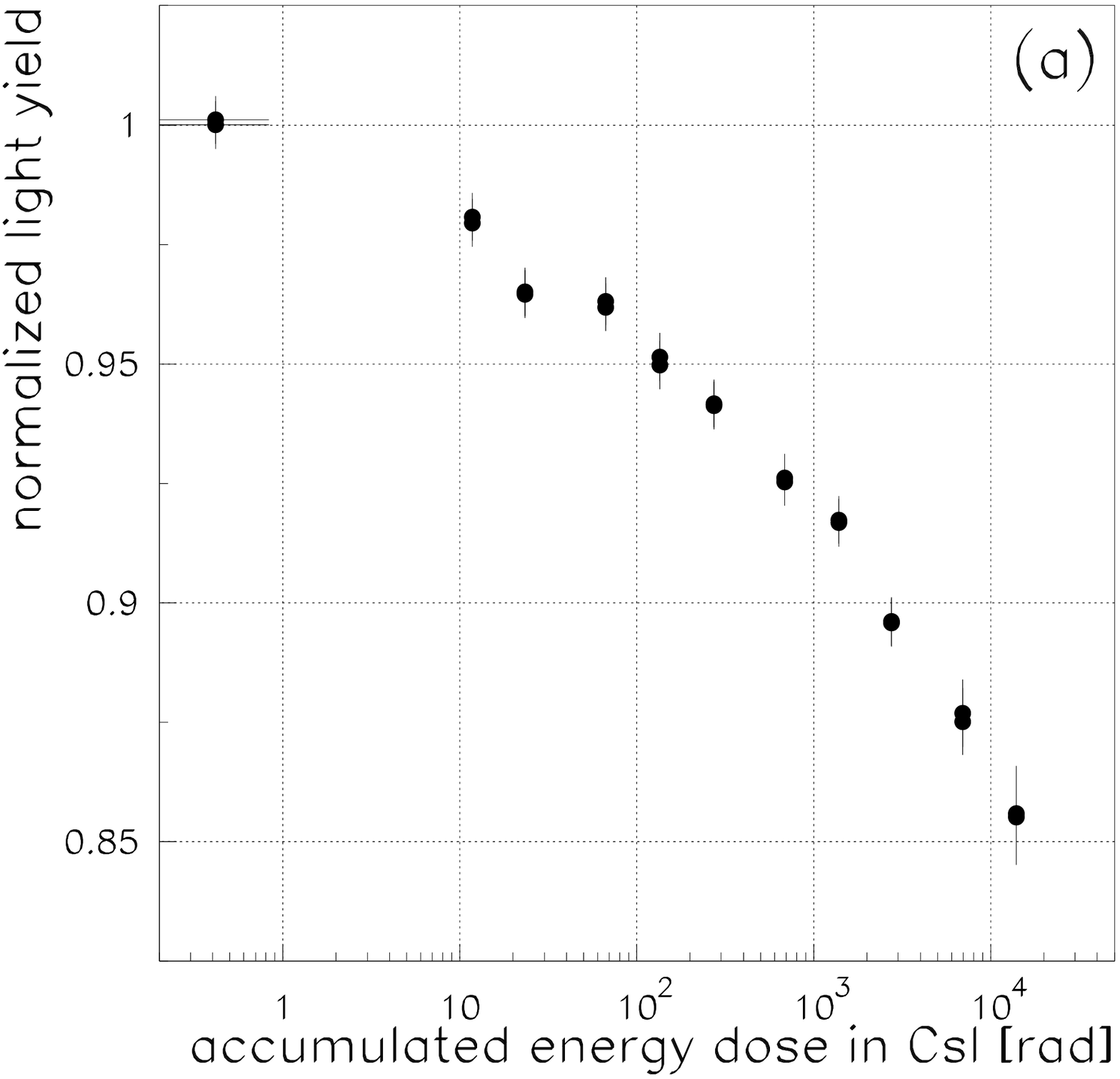,width=0.48\textwidth}} &
\mbox{\epsfig{file=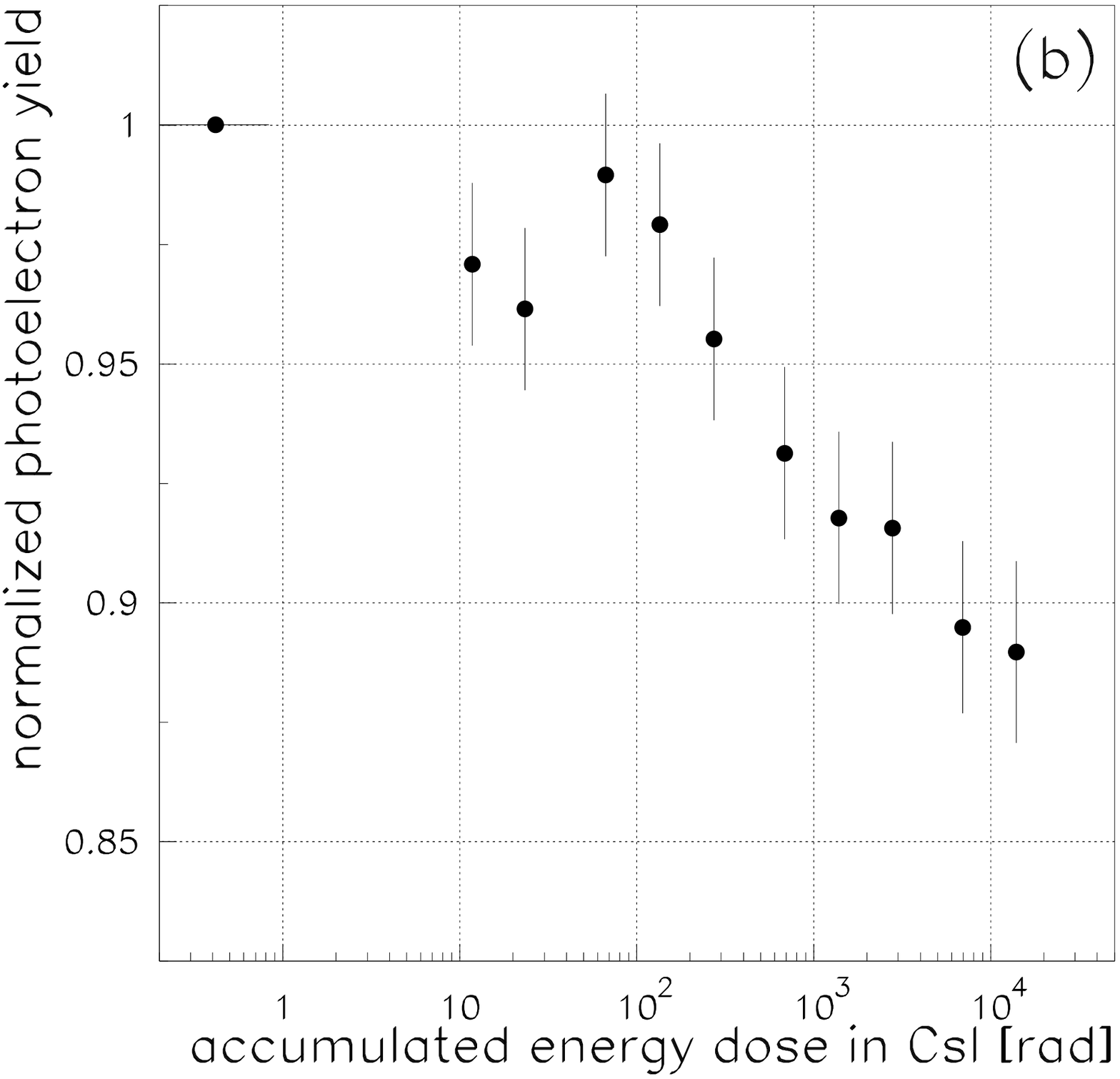,width=0.48\textwidth}} \\
\end{tabular}
\caption[]{Decrease of crystal light yield for photomultiplier (a) and 
photodiode (b) readout measured four days after each irradiation, 
normalized to the yield before irradiation vs. accumulated energy dose in CsI.}
\label{fig:rad} 
\end{figure}

The influence of the accumulated energy dose on the crystal light yield was
investigated with photomultiplier (\cs\ source) and photodiode (\y\ source)
readout. Directly after irradiation, a decrease of the crystal light yield is 
observed which reaches a constant level a few hours (days) after irradiation 
with low (higher) doses.
Therefore, the light output on the fourth day after irradiation was chosen as 
a measure  for the reduction of light yield after successive irradiations 
and plotted versus the accumulated energy dose in CsI in
Fig.~\ref{fig:rad}. In the case of photodiode readout, the diodes were 
fixed with a minimal airgap at the crystal rear, in order to prevent radiation
damage to the diodes and to allow for alternating photomultiplier and 
photodiode measurements of the same crystal. Therefore, the measured \py\ is 
smaller than with glued diodes. The errors are larger because of the 
reproducibility of the optical coupling.
At an accumulated energy dose of $\sim 14\krad$, light reductions of  
$(14.5 \pm 1.0)\%$ and $(11 \pm 3)\%$ are measured with the photomultiplier 
and photodiode readout setup, respectively. In repeated measurements during 
the week between two irradiations and months after the last irradiation no 
recovery of the light output was observed. On the other hand, the uniformity 
curve of the crystal shows no change, even at the highest dose.\\

\begin{figure}[t]
\begin{center}
\unitlength=1mm
\begin{picture}(80,80)
\put(24,62){\makebox(0,0)[cc]{$^{137}$Cs}}
\put(24,56){\makebox(0,0)[cc]{$|$}}
\put(35.5,55){\makebox(0,0)[cc]{$^{40}$K}}
\put(35.5,49){\makebox(0,0)[cc]{$|$}}
\put(67,68){\makebox(0,0)[cc]{LED}}
\put(67,62){\makebox(0,0)[cc]{$|$}}
\mbox{\epsfig{file=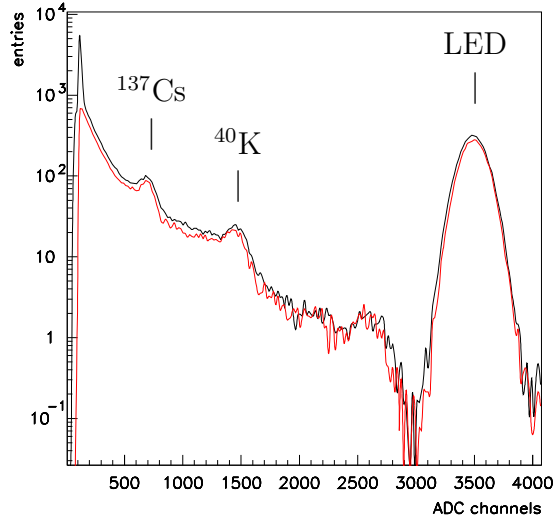,width=0.6\textwidth}} 
\end{picture}
\caption[]{Pulse height spectrum with additional background caused by the 
phosphorescence. The dashed curve corresponds to the same spectrum measured 
some days later showing the strong reduction of phosphorescence light at 
ADC channel $\sim 200$.}
\label{fig:glow} 
\end{center}
\end{figure}

The observed phosphorescence contributes to an additional background near 
threshold as shown in Fig.~\ref{fig:glow}. It also has an effect on the width
of the $662\keV$ photon peak from the \cs\ source, which is used as a measure 
of the crystal light yield. If one quadratically subtracts the width of the 
\cs\ peak without afterglow from the corresponding one with afterglow 
(measured a few hours after irradiating with the highest applied dose) one 
gets a contribution of the afterglow to the width of $(30\pm 5)\keV$. This 
corresponds to a change in energy resolution $\sigma_E/E$ from 
$6.6\%$ to $8.0\%$ measured with the photomultiplier setup.
The observed reduction of crystal light yield is similar to results 
of~\cite{ref:kazui}, which observe typical decreases of $5\%$, $15\%$, and
$30\%$ at $10\rad$, $100\rad$, and $1\krad$, respectively, without uniformity 
changes for crystals from the same manufacturer. 
Older studies~\cite{ref:blucher,ref:oldrad1} often found higher 
decreases, up to $25\%$ below $\sim 100\rad$ and $50\%$ at 
$\sim 10\krad$, probably due to higher levels of impurities in 
crystals~\cite{ref:oldrad2}. All these studies used coarser steps in the 
applied doses.

%%%%%%%%%%%%%%%%%%%%%%%%%%%%%%%%%%%%%%%%%%%%%%%%%%%%%%%%%%%%%%%%%%%%%%%%%%%%%

\section{Conclusions}
\label{sec:concl}

A photomultiplier setup in conjunction with a temperature stabilized reference
system was developed and put into operation. It allows precise measurements
of \CsI\ crystal light yield and uniformity with an accuracy of $0.3\%$. This
setup was used for optimization of the crystal light yield and for
tuning the light output uniformity. An optimized coupling of Silicon PIN
photodiodes to crystals was established with respect to \py\ and electronic 
noise. For long \CsI\ crystals of 19.4 radiation lengths, \py s of 
$\sim 6000\emi/\MeV$ for wavelength shifter readout and $\sim 8500\emi/\MeV$
for direct readout and corresponding equivalent noise energies of 
$\sim 100\keV$ (RMS) were achieved. The temperature dependence of the relative 
crystal \ly\ and the \py\ were determined to be $+(0.40\pm 0.01)\%/\K$ and
$+(0.28\pm 0.02)\%/\K$ for photomultiplier and photodiode readout, 
respectively.
After irradiation of one crystal with an intense $\co$ $\gamma$ source in
steps up to $\sim 14\krad$ accumulated energy dose, a reduction in
crystal light yield in the range of $10\%$~to~$15\%$ and afterglow effects 
caused by phosphorescence, were observed.

%%%%%%%%%%%%%%%%%%%%%%%%%%%%%%%%%%%%%%%%%%%%%%%%%%%%%%%%%%%%%%%%%%%%%%%%%%%%%

\section*{Acknowledgements}

We would like to express our thanks to Drs. P.~Eckstein, R.~Schwierz, and 
R.~Waldi for their advice and support and to S.~Jugelt for his help in the
measurements. Especially we wish to thank H.~Futterschneider, R.~Krause, and
F.~M\"oller for their help setting up the experiments and fighting the 
electronics noise. 
Valuable comments and discussions with the members of the BABAR Calorimeter 
Group, especially Drs.~C.~Jessop, R.~Wang, and M.~Pertsova, are greatly 
appreciated.
Special thanks go to Dr. K. Prokert for his competent help during 
all phases of the radiation damage sudies and their dosimetry.

%%%%%%%%%%%%%%%%%%%%%%%%%%%%%%%%%%%%%%%%%%%%%%%%%%%%%%%%%%%%%%%%%%%%%%%%%%%%%
%%%%%%%%%%%%%%%%%%%%%%%%%%%%%%%%%%%%%%%%%%%%%%%%%%%%%%%%%%%%%%%%%%%%%%%%%%%%%


\begin{thebibliography}{99}

\bibitem{ref:cleo}
{C. Bebek (CLEO)},
\newblock  Nucl. Instr. Meth. A265 (1988) 258

\bibitem{ref:cb}
{E. Aker et al. (Crystal Barrel)},
\newblock  Nucl. Instr. Meth. A321 (1992) 69

\bibitem{ref:babar}
{BABAR Collaboration, D. Boutigny et al.},
\newblock  Technical Design Report, SLAC-R-95-457 

\bibitem{ref:belle}
{BELLE Collaboration, M. T. Cheng et al.},
\newblock  Technical Design Report, KEK 95-01

\bibitem{ref:grassmann}
{M. Grassmann, E. Lorenz, H.-G. Moser},
\newblock Nucl. Instr. Meth. A228 (1985) 323

\bibitem{ref:blucher}
{E. Blucher et al. (CLEO)},
\newblock  Nucl. Instr. Meth. A249 (1986) 201

\bibitem{ref:valentine}
{J. D. Valentine et al.},
\newblock IEEE trans. Nucl. Sci. 40 (1993) 1267

\bibitem{ref:hamamatsupm}
{HAMAMATSU Corp.},
\newblock  Photomultiplier Tubes, Hamamatsu Photonics K.K., 
Electron Tube Center (1994), Cat. No. TPMO 0002E02, Apr. 94

\bibitem{ref:bn206}
{M. King et al.},
\newblock  A study on crystal wrapping, BABAR Note 206, 1995

\bibitem{ref:bn175}
{J. Brose et al.},
\newblock  CsI crystal uniformity specification and quality control, 
BABAR Note 175, 1995

\bibitem{ref:hamamatsupd}
{HAMAMATSU Corp.},
\newblock  Si PIN Photodiodes S3590- / S2744- / S3204- / S3584- / S3588-08
(preliminary data), 
Hamamatsu Photonics K.K. Solid State Division (1994), Cat. No. KPIN 1023E01, Oct. 95;\\
{HAMAMATSU Corp.},
\newblock  Photodiodes, Hamamatsu Photonics K.K. Solid State Division
(1994), Cat. No. KPD 0001E02, Feb. 94

\bibitem{ref:bn216}
{C. Jessop et al.},
\newblock  Development of the front end readout for the BABAR CsI(Tl) 
calorimeter, BABAR Note 216, 1995

\bibitem{ref:bn242}
{J. Brose et al.},
\newblock  Optimization of photodiode readout of CsI(Tl) crystals,
BABAR Note 242, 1995

\bibitem{ref:bn241}
{G. Dahlinger},
\newblock  Wrapping and tuning studies for large CsI crystals,
BABAR Note 241, 1996

\bibitem{ref:bn236}
{C. Jessop, J. Harris},
\newblock  Performance test of Hamamatsu 2744-08 diodes for the BABAR 
calorimeter front end readout and proposal for reliability issues, 
BABAR Note 236, 1995

\bibitem{ref:bn270}
{C. Jessop},
\newblock  Development of direct readout for CsI Calorimeter,
BABAR Note 270, 1995

\bibitem{ref:kobayashi}
{M. Kobayashi, P. Carlson, S. Berglund},
\newblock Nucl. Instr. Meth. A281 (1989) 192

\bibitem{ref:bn355}
{G. Dahlinger, J. Brose},
\newblock  Temperature dependence and radiation hardness of CsI(Tl) crystals,
BABAR Note 355, 1997

\bibitem{ref:dose} 
{J.H. Hubbell, S.M. Seltzer},
\newblock
Tables of X-Ray Mass Attenuation Coefficients and Mass Energy-Absorption 
Coefficients $1\keV$ to $20\MeV$ for Elements Z=1 to 92 and 48 Additional 
Substances of Dosimetric Interest,\\ 
\mbox{NISTIR 5632-Web Version}\\   
http://physics.nist.gov/PhysRefData/XrayMassCoef/cover.html

\bibitem{ref:kazui}
{K. Kazui et al.},
\newblock Nucl. Instr. Meth. A394 (1997) 46

\bibitem{ref:oldrad1}
{S. Schl\"ogl, H. Spitzer, K. Wittenburg},
\newblock Nucl. Instr. Meth. A242 (1985) 89

\bibitem{ref:oldrad2}
{M. Kobayashi, S. Sakuragi},
\newblock Nucl. Instr. Meth. A254 (1987) 275


\end{thebibliography}
\end{document}